\documentclass[natbib209]{emulateapj}

\usepackage{amsmath}
\usepackage{epsfig}

\newcommand{\pd}{\partial}

\newcommand{\bdot}{\mbox{\boldmath $\cdot$}}
\newcommand{\del}{\mbox{\boldmath $\nabla$}}
\newcommand{\curl}{\mbox{\boldmath $\nabla \times$}}
\newcommand{\dv}{\mbox{\boldmath $\nabla \bdot$}}
\newcommand{\cross}{\mbox{\boldmath $\times$}}

\newcommand{\vv}{{\bf v}}

\newcommand{\rh}{\overline{\rho}}

\newcommand{\uvr}{\mbox{\boldmath $\hat{r}$}}

\newcommand{\uvp}{\mbox{\boldmath $\hat{\phi}$}}

\def\del{\nabla}
\def\cross{\times}
\def\avg{\bar}
\def\vec{\boldsymbol}
\def\scrD{\mathcal{D}}
\def\scrR{\mathcal{R}}

\def\threed{3-D }

\shorttitle{Meridional Circulation in Stars}

\shortauthors{Featherstone \& Miesch}

\begin{document}

\title{Meridional Circulation in Solar and Stellar Convection Zones}

\author{Nicholas A. Featherstone}
\affil{JILA, University of Colorado, Boulder, CO 80309-0440}
\email{feathern@solarz.colorado.edu}
\author{Mark S. Miesch}
\affil{High Altitude Observatory, National Center for Atmospheric Research, Boulder, CO, 80307-3000, USA: miesch@ucar.edu}

\begin{abstract}
We present a series of 3-D nonlinear simulations of solar-like convection, carried out using the Anelastic Spherical Harmonic (ASH) code, that are designed to isolate those processes that drive and shape meridional circulations within stellar convection zones.  These simulations have been constructed so as to span the transition between solar-like differential rotation (fast equator/slow poles) and ``anti-solar' differential rotation (slow equator/fast poles). Solar-like states of differential rotation, arising when convection is rotationally constrained, are characterized by a very different convective Reynolds stress than anti-solar regimes, wherein convection only weakly senses the Coriolis force.  We find that the angular momentum transport by convective Reynolds stress plays a central role in establishing the meridional flow profiles in these simulations. We find that the transition from single-celled to multi-celled meridional circulation profiles in strong and weak regimes of rotational constraint is linked to a change in the convective Reynolds stress, a clear demonstration of gyroscopic pumping.  Latitudinal thermal variations differ between these different regimes, with those in the solar-like regime conspiring to suppress a single cell of meridional circulation, whereas the cool poles and warm equator established in the anti-solar states tend to promote single-celled circulations.  Though the convective angular momentum transport becomes radially inward at mid-latitudes in anti-solar regimes, it is the meridional circulation that is primarily responsible for establishing a rapidly-rotating pole.  We conclude with a discussion of how these results relate to the Sun, and suggest that the Sun may lie near the transition between rapidly-rotating and slowly-rotating regimes.
\end{abstract}

\section{Introduction}\label{sec:intro}

Mean (longitudinally-averaged) flows play an essential role in all recent models
of the solar activity cycle.  Differential rotation (the mean zonal flow) 
generates toroidal field from poloidal field and amplifies it, tapping the
kinetic energy of shear.  This is the well-known $\Omega$-effect and it is
generally held responsible for the observed disparity between the toroidal 
flux inferred from bipolar active regions and the much smaller poloidal
flux inferred from high-latitude magnetograms.

The role of the mean meridional circulation (MC) in the operation of
the solar dynamo is more controversial but potentially no less profound.
As discussed in \S\ref{sec:blft}, mean-field dynamo models have shown that
the amplitude and structure of the MC can have a substantial influence
on cycle properties, and in particular, it may largely regulate the
cycle period.  

Yet, despite the central role that mean flows play in solar and
stellar dynamo models, their existence is often taken for granted.
Most mean-field dynamo models published in the literature are
kinematic in the sense that the velocity field is specified so the
origin of mean flows lies outside the scope of the model.
Non-kinematic mean-field models do exist but many take a heuristic
approach, introducing parameterizations for convective momentum and
energy transport that produce mean flows with preconceived amplitudes
and profiles \cite[e.g.][]{rempe06}.

The utility of these kinematic and heuristic approaches to dynamo
modeling rests on the availability of sufficient observational data
to constrain the mean flows that are employed.  However, this
observational data is limited.  Global helioseismic inversions provide
a reliable measure of the time-averaged angular velocity profile
$\Omega(r,\theta)$ throughout most of the solar convection zone
\citep[CZ,][]{thomp03,howe09}, but as discussed in \S\ref{sec:obs}
relatively little is known about the MC.  Even less is known about
mean flows in other stars (see \S\ref{sec:sas}).  Yet, stars provide
an invaluable test of dynamo models; any viable model of the solar
activity cycle must also account for cyclic magnetic activity in
solar-like stars.

Without some theoretical guidance as to what meridional and zonal
flows one should expect to find in solar and stellar convection zones,
kinematic dynamo models cannot completely account for observed
patterns in solar and stellar magnetic activity.  Furthermore,
theoretical guidance is needed to help motivate and interpret ongoing
observational investigations of the solar meridional circulation
(\S{\ref{sec:obs}) which seek to probe more generally the internal
dynamics of the Sun and the origins of solar magnetic activity.

In a previous paper, \citep[][hereafter Paper 1]{miesc12b}, we sought
such theoretical guidance based only on the observed 
structure of solar mean flows and robust dynamical balances obtained from the
fundamental magnetohydrodynamic (MHD) equations.  From this alone we
were able to identify gyroscopic pumping as the principle physical
mechanism that maintains the solar MC.  This work is summarized in
\S\ref{sec:background}.  However, further progress requires an understanding
of turbulent solar convection and in particular the convective 
angular momentum transport.  To this end, in this paper, we have 
initiated a series of global convection simulations designed to 
elucidate the nature of convective transport in rotating, stratified,
spherical shells and the mean flows it produces.  
We focus here on solar-like stars with convective
envelopes but similar dynamics are likely to occur in other types
of stars with different convection zone geometries.

In \S\ref{sec:background} we survey how this 
series of simulations fits within the context of previous theoretical
and observational work and in \S\ref{sec:experiments} we discuss
the motivation and methodology behind how the numerical experiments
were set up.  We then present
an overview of our results in \S\ref{sec:regimes}, identifying two
distinct dynamical regimes that exhibit qualitatively different mean
flows.  As in previous studies (see \S\ref{sec:models}), 
the dynamical regimes are distinguished by the degree of
rotational influence on the convection, as quantified by the Rossby
number $Ro = U / (2\Omega_0 L)$, where $U$ and $L$ are
characteristic convective velocity and length scales and $\Omega_0$ is 
mean rotation rate.  Fast rotators produce solar-like $\Omega$
profiles (fast equator, slow poles) with multi-celled MC profiles
while slow rotators produce anti-solar $\Omega$ profiles
(slow equator, fast poles) with single-celled MC profiles (poleward
flow in the upper CZ and equatorward in the lower CZ).  We demonstrate
that these slow and fast regimes can be achieved not only by changing
the rotation rate but also by changing the mechanical and thermal
dissipation, which in turn impacts the buoyancy driving.  In
\S\ref{sec:interp} we interpret the resulting MC profiles as a consequence
of the changing nature of the convective Reynolds stress, within the
context of gyroscopic pumping.  We address the establishment and
significance of thermal gradients in \S\ref{sec:heat}.

Which regime is the Sun in?  Since it has a solar-like $\Omega$
profile by definition, one would expect it to be classified as a rapid
rotator.  However, the situation is more subtle than this.  In
\S\ref{sec:sas} we demonstrate that the transition between solar and
anti-solar differential rotation is intimately linked to the MC and
may be sensitive to the complex dynamics occurring in the boundary
layers, namely the tachocline and the near-surface shear layer.  The
Sun lies very close to this transition, which may imply that it has
somewhat atypical mean flow profiles.  This in turn has important
implications for solar dynamo models and their application to other
stars.  These issues too are addressed in \S\ref{sec:sas} and our
principle results and conclusions are summarized in
\S\ref{sec:summary}.

\section{Theoretical and Observational Background}
\label{sec:background}
\subsection{Meridional Circulation and the Solar Dynamo}\label{sec:blft}

The potential significance of the MC to the operation of the solar
dynamo is apparent from the observed evolution of the radial magnetic
flux in the solar photosphere.  In particular, the cyclic polarity
reversals of polar fields observed at the solar surface are often
attributed to the poleward transport of the trailing magnetic flux in
emerging active regions due to the combined action of the meridional
flow and turbulent diffusion \citep{sheel05,bauma06,schri08}.  
Though the timing of the most recent north polar reversal in 2012
challenges this paradigm \citep{shiot12}, it is clear that 
the meridional flow plays an important role in redistributing
the magnetic flux that threads through the solar surface.

These observations and models of surface flux transport help to
justify the prominent role that the MC plays in many recent dynamo
models of the solar cycle.  This role is most dramatic in a class of
models known as flux-transport (FT) dynamo models, in which the
equatorward advection of toroidal flux by the meridional flow in the
CZ and tachocline accounts for the observed
migration of sunspots toward the equator over the course of a cycle,
known as the solar butterfly diagram
\citep{wang91,dikpa99b,kuker01,dikpa06,yeate08,munoz09,jouve07}.  By
regulating the solar butterfly diagram, the MC also regulates the
cycle period.  

In a subset of flux-transport dynamo models that operate in the
so-called advection-dominated regime, the MC plays another important
role.  In practice, most existing advection-dominated flux-transport
(FT) dynamo models are also Babcock-Leighton (BL) dynamo models in
which the main source of poloidal flux lies in the buoyant
destabilization, emergence and subsequent dispersal of toroidal flux
\citep{dikpa09,charb10}.  Mean-field parameterizations of the BL mechanism typically
take the form of a poloidal source term confined to the surface layers
\cite[e.g.]{dikpa99b,rempe06,munoz10}.  Thus, in order for the dynamo to operate, 
there must be a coupling mechanism that transports magnetic flux from the 
poloidal source region near the surface to the region where the toroidal 
field is generated, which generally lies in the lower CZ.  In
advection-dominated BL/FT dynamo models, this coupling mechanism is
the MC.

Changes in the amplitude and structure of the meridional flow in FT
dynamo models can have a dramatic effect on the length and amplitude
of magnetic cycles.  Faster meridional flows or multiple cells in
latitude can shorten the cycle period and influence the polar field
strength while slower flows can lengthen cycles and suppress activity,
potentially inducing a grand minimum
\citep{dikpa99b,jouve07,schri08,jiang09,karak10}.
Multiple cells in radius can have an even greater impact, 
disrupting the operation of the dynamo and potentially reversing
the sense of the butterfly diagram \citep{jouve07}.

The profound role of the MC in contemporary solar dynamo models has
motivated intense, ongoing observational efforts to probe the meridional
flow structure in the deep CZ.  However, this is a formidable challenge,
largely because the flow is expected to be so weak; typical amplitudes
used in FT dynamo models are only 2-3 m s$^{-1}$ near the base of the
CZ.  Thus, observational constraints are currently limited but may
improve with more data, as we now address.

\subsection{Observational  Context}\label{sec:obs}

One of the most notable triumphs of helioseismology has been the
mapping of the internal solar rotation profile $\Omega$ as a function
of radius $r$ and colatitude $\theta$ \citep{thomp03,howe09}.  These
helioseismic rotation inversions span most of the solar convection
zone, though they become less reliable near the poles and in the deep
interior due both to the coarser resolution of the inversion kernels
and the smaller amplitude of the intrinsic frequency splitting near
the rotation axis.  They reveal a monotonic decrease of $\Omega$ by
about 30\% from equator to pole that persists throughout most of the
CZ, with little radial variation apart from two boundary layers near
the bottom and top, known as the tachocline and the near-surface shear
layer respectively.

The most well-established result for the solar MC, meanwhile, is the
existence of a systematic poleward flow of 10-20 m s$^{-1}$ at the
surface at latitudes below about $\pm $ 50-60$^\circ$.  However,
different techniques often give different results as to the subsurface
structure, the high-latitude structure, and even the low-latitude
amplitude.  For example, measurements of the meridional flow based on
the correlation tracking of magnetic features generally give somewhat
lower flow speeds than those based on Doppler shifts and local
helioseismic inversions \citep{ulric10}.  \cite{dikpa10} attribute this
difference to the effects of turbulent diffusion.  Much of the
variation in amplitude estimates arises from the intrinsic variability
of the flow itself, which is known to change with the solar cycle
\citep{ulric10,basu10,hatha10,hatha11}.  However, these solar cycle
variations can be understood as perturbations about a persistent
background circulation (poleward near the surface at low-mid
latitudes) that are induced by surface magnetism and its impact on
radiative cooling.  In this paper we are interested only in the
persistent background circulation so we disregard the solar cycle
variations.

At higher latitudes the results are less clear.  The recent Doppler
analysis by \cite{ulric10} suggests the possible presence of a
high-latitude counter-cell with equatorward surface flow at latitudes
above about 60$^\circ$.  This feature is most apparent near solar
minimum and may be intermittent.  Indeed, \cite{dikpa10b} argue that
the absence of such a high-latitude counter cell through much of solar
cycle 23 may account for the relatively deep minimum of 2008-2009 and
the delayed onset of solar cycle 24.  However, evidence for such a
high-latitude counter-cell based on feature tracking and helioseismic
inversions has so far been lacking, with the former suggesting that
poleward flow may persist all the way to the poles \citep{right12} and
the latter being thus far inconclusive \citep{zhao12}.  Further
insight into this question should be possible in the coming years with
the availability of longer time series from the Helioseismic Magnetic
Imager (HMI) on the Solar Dynamics Observatory (SDO), which has the
higher spatial resolution needed to minimize limb effects, and also
with the advent of future missions such as Solar Orbiter, which will
have a high-latitude vantage point.

Of these three principle techniques for measuring the meridional flow
(Doppler measurements, feature tracking, and local helioseismic
inversions), only helioseismic inversions and feature tracking can
probe the meridional flow below the surface. While the observational
consensus is that there is a systematic poleward flow of about 10-20 m s$^{-1}$ at
latitudes below about 60$^\circ$ in the near-surface layers of the Sun, the subsurface structure is more controversial.  The first studies based on SOHO/MDI data by
\citet{giles97} suggested that the poleward flow extended down to at least 0.96$R$.
Subsequent studies suggested that this poleward flow may extend significantly 
deeper, spanning at least the upper half of the CZ \citep{braun98,chou01}.  However,
more recent observations and inversions based on longer time series, new analysis 
techniques, and new data from HMI/SDO have begun to cast doubt on this conclusion.  

\citet{hatha12b} argued based on surface feature tracking that the return equatorward 
flow required for mass conservation may occur as shallow as 0.93$R$.   Inferring subsurface
flows with feature tracking relies on the hypothesis that larger
horizontal scales in the photosphere trace convective structures that
extend deeper into the CZ and may thus be advected by deeper mean
flows \citep{hatha12}.  \cite{hatha12,hatha12b} tests this hypothesis
by comparing inferred zonal flows to helioseismic rotational
inversions and then applies it to infer the meridional flow.  Results
suggest a flow reversal from poleward to equatorward at a depth of
roughly 50 Mm ($r \sim $ 0.93, where $R$ is the solar radius).  This
is notably shallow compared to earlier estimates obtained from local
helioseismic inversions.

More recent helioseismic investigations,
fueled by the high-quality data from HMI/SDO and by the availability
of long time series from GONG and SOHO/MDI\footnote{the Global Oscillations
Network Group (GONG) and the Michelson Doppler Imager (MDI) on 
the Solar and Heliospheric Observatory (SOHO)}, suggest that this
picture may be changing.  Time-distance helioseismic inversions by \citet{zhao13}, taking into account systematic center-to-limb 
effects on the inversions, appear to support this conclusion, showing evidence for multiple 
reversals in the latitudinal flow component, the shallowest of which being 
0.91 $R$ \citep[see also][]{mitra07}.  However, such a shallow reversal is not 
found in global inversions of low-order oscillation modes where the meridional flow 
enters as higher-order effect \citep{schad12}.

Given the high stakes from the perspective of solar and stellar dynamo
theory, the current uncertainty with regard to the meridional flow
structure in the deep convection zone, and the vibrancy and promise of
ongoing observational efforts, further theoretical insight is sorely
needed.  Here we seek such insight from convection simulations,
after a brief overview of the fundamental processes involved
(\S\ref{sec:background}).

\subsection{Dynamical Models}\label{sec:models}

As noted above, kinematic mean-field dynamo models cannot provide any insight
into the origin and structure of the MC.   For this, dynamical models are
required.  Such models must address the transport of energy and momentum
by convection and the subsequent nonlinear feedbacks involving the MC, 
differential rotation, and thermal stratification \citep{miesc09}.  This
can be done either explicitly through global 3-D convection simulations
or implicitly through non-kinematic mean-field models that include
parameterizations for the turbulent transport.

Global convection simulations have revealed two distinct dynamical
regimes characterized by qualitatively different differential rotation 
(DR) profiles \citep{gilma76,gilma77,glatz82,aurno07,brun09,matt11,gasti13,guerr13,gasti14,kapyl14}.  
At strong rotational influence $Ro \lesssim 1$ the differential rotation 
is ``solar-like'' in the sense that the equator spins faster than the
poles. In other words $\Delta \Omega \equiv \Omega_{eq}-\Omega_p > 0$
where $\Omega_{eq}$ is the equatorial rotation rate at the outer 
boundary of the domain, and $\Omega_p$ is some measure of the 
rotation rate at high latitudes, poleward of $70^\circ$.  For the
weakly rotating regime, $Ro \gtrsim 1$, the differential 
rotation is referred to as ``anti-solar'', meaning $\Delta \Omega < 0$.
The anti-solar regime has often been attributed to the tendency
for the convection to conserve angular momentum locally, 
spinning up as motions approach the rotation axis
\citep{gilma77,aurno07}.

Though these regimes can be achieved by varying the rotation rate
$\Omega$ (or its non-dimensional equivalent), several studies have
demonstrated that the transition from solar to anti-solar differential
rotation can also be achieved by keeping $\Omega$ (or its non-dimensional
equivalent) fixed and changing the convective driving through the Rayleigh
number \citep{gasti14,kapyl14}.  This tends to increase $Ro$ by
decreasing the convective turnover time $L/U$.  In other words,
convective velocity scales increase while length scales decrease.
Near the solar/anti-solar transition, \cite{gasti14} and \citet{kapyl14} 
also find that the DR exhibits bistability, with both solar and anti-solar 
regimes possible depending on initial conditions and hysteresis.  

Many of these studies reported a qualitative change in the MC profile 
in the two regimes \citep{matt11,guerr13,gasti13,gasti14,kapyl14}.  
The anti-solar regime ($Ro < 1$) often exhibits a single circulation cell
per hemisphere, with poleward flow in the upper CZ and equatorward 
flow in the lower CZ.  Meanwhile, the solar regime ($Ro > 1$) often
exhibits multiple cells in latitude and radius, aligned with the 
rotation axis outside the tangent cylinder\footnote{The tangent
cylinder is a cylindrical surface parallel to the rotation axis and
tangent to the base of the CZ}.

Recent high-resolution models have also begun to capture aspects 
of the solar near-surface shear layer (NSSL) wherein small-scale convection in the
surface layers establishes an inward $\Omega$ gradient \citep{guerr13b,hotta14b}.
As suggested by \cite{miesc11}, the NSSL in these simulations is established
by a radially inward angular momentum transport, is maintained by meridional
Reynolds stress, and is intimately linked to a poleward meridional flow.
This will be discussed further in \S\ref{sec:nssl} below.

The solar and anti-solar regimes in global convection simulations stand in stark 
contrast to mean-field models which typically produce solar-like DR profiles 
\cite[e.g.][]{kuker11,kitch12,kitch12b}.  Anti-solar profiles
generally only occur if there is a strong, baroclinic, single-celled 
meridional flow induced by additional factors such as polar starspots 
or tidal forcing from a binary companion \citep{kitch04,rudig07}.  

Mean-field models of solar-like stars generally exhibit single-celled
MC profiles but their origin and structure vary depending on the modeling
approach.  One approach is to seek steady-state solutions of the mean-field 
equations with heuristic or model-based prescriptions for the convective 
Reynolds stress and heat transport.  In such models the MC profile is 
largely determined by departures from thermal wind balance due to the
meridional Reynolds stress, which is typically modeled as a turbulent
diffusion \citep{kitch12,dikpa14}.  This can lead to
nonlocal driving of the MC from the upper and lower boundary layers \citep{kitch12}
or to high-latitude counter-cells for steep density stratifications
\citep{dikpa14}.  However, in this case, the form of the MC profile is 
sensitive to the nature of the turbulent viscosity profile that is imposed. 
This alone breaks the degeneracy of the thermal wind balance equation with 
respect to the meridional flow (see \S\ref{sec:gp} below).

When the full time-dependent mean-field equations are solved, the driving 
of the meridional flow is fundamentally different.  Here the MC profile
is determined mainly by the non-diffusive component of the convective 
angular momentum transport, which is typically modeled by the $\Lambda$-effect
in the zonal momentum equation \citep{rempe05,rempe06b}.  This is the process
of gyroscopic pumping that will be described in detail in \S\ref{sec:gp}
and that is responsible for the establishment of MC in our convection
simulations, as demonstrated in \S\ref{sec:amom}.

%>>>>>>>>>>>>>>>>>>>>>>>>>>>>>>>>>>>>>>>>>>>>>>>>>>>>>>>>>>>>>>>

\section{Description of Numerical Convection Experiments}\label{sec:experiments}

We present a series of numerical experiments designed to examine solar-like convection and its resulting mean flows under conditions that span a range of rotational constraints and convective efficiency.   We model the solar convection zone using the anelastic approximation (Gough 1969; Gilman \& Glatzmaier 1980).  Such an approach is appropriate in deep stellar interiors where plasma motions are subsonic and perturbations to thermodynamic variables are small compared to their mean, horizontally averaged values.  As thermodynamic perturbations are small, thermodynamic variables are linearized about a spherically symmetric reference state with density $\avg{\rho}$, pressure $\avg{P}$, temperature $\avg{T}$. and specific entropy $\avg{S}$.  Fluctuations about this state are denoted as $\rho$, P, T, and S.  In the uniformly-rotating reference frame of the star, the anelastic equations under these assumptions may be expressed as

\begin{equation}
  \label{eq:div mass flux}
  \del \cdot(\avg{\rho}\vec{v}) = 0,
\end{equation}

\begin{equation}
  \label{eq:momentum}
\begin{split}
  \avg{\rho}\left[ \frac{D\vec{v}}{Dt} +
    2 \vec{\Omega}_0 \cross \vec{v} \right]
  =
  -\del P + \rho \vec{g} - \del \cdot \scrD,
\end{split}
\end{equation}
and

\begin{equation}
  \label{eq:entropytwo}
  \begin{split}
  \avg{\rho}\avg{T}\frac{DS}{Dt} =&
  \del \cdot [\kappa \avg{\rho} \avg{T} \del S +\kappa_r \avg{\rho} c_p \del(\avg{T}+T)]\\
                    &+ 2 \avg{\rho}\nu \left[e_{ij}e_{ij} - \frac{1}{3}(\del \cdot
\vec{v})^2\right]. 
  \end{split}
\end{equation}

The velocity $\vec{v}$ expressed in spherical coordinates is $\vec{v} = (v_r,v_{\theta},v_{\phi})$ relative to a frame rotating at constant angular velocity $\mathbf{\Omega_o}$, $\vec{g}$ is the gravitational acceleration, and $\scrD$ is the viscous stress tensor given by 

\begin{equation}
  \scrD_{ij} = -2 \avg{\rho} \nu \left[e_{ij}
    - \frac{1}{3}(\del \cdot \vec{v})\delta_{ij} \right],
\end{equation}
where $e{_{ij}}$ is the strain rate tensor.  The kinematic viscosity is denoted by $\nu$ and the thermal diffusivity by $\kappa$. The radiative diffusivity is indicated by $\kappa_r$, $c_p$ is the specific heat at constant pressure, and $D/Dt$ is the advective derivative.  $\kappa_r$ and $c_p$ remain constant in time.  

This set of equations is closed by assuming the thermodynamic fluctuations satisfy the linear relations
\begin{equation}
  \frac{\rho}{\avg{\rho}} = \frac{P}{\avg{P}} - \frac{T}{\avg{T}}
    =  \frac{P}{\gamma \avg{P}} - \frac{S}{c_p},
\end{equation}
assuming the ideal gas law
\begin{equation}
  \avg{P} = \scrR \avg{\rho} \avg{T},
\end{equation}
where $\scrR$ is the gas constant.  

We evolve the anelastic equations using the anelastic spherical harmonic (ASH) code, details of which may be found in Clune et al. (1999) and Brun et al. (2004).  ASH solves the \threed MHD equations in a rotating, spherical shell using a pseudo-spectral approach.  Spherical harmonics are employed in the horizontal dimension, and 4th-order finite-differences in the radial direction.  ASH employs a poloidal/toroidal decomposition for the mass flux vector to satisfy the divergence-free constraint of eq. \ref{eq:div mass flux}, with

\begin{equation}
	\avg{\rho}\vec{v} = \del\cross\del\cross(W\vec{e}_r)+\del\cross(Z\vec{e}_r).
\end{equation}

Here $W$ and $Z$ are the poloidal and toroidal streamfunctions respectively, and are the quantities that are explicitly evolved in time (as opposed to the individual components of $\vec{v}$).  The unit vector in the radial direction is indicated by $\vec{e}_r$.

The suite of convection zone models that we examine are identical in all respects save for the diffusivities and the rotation rate.  The reference state temperature, pressure, and functional form of $\kappa_r$ are derived from a one-dimensional solar structure model \citep{brun02b}.  While the specific heat $c_p$ of the one-dimensional structure model varies weakly throughout the solar convection zone, we have adopted a constant value of $c_p$, taken from the base of that model.  This requires us to iteratively solve for the reference state density $\avg{\rho}$, using the 1-D model density as an initial guess, until a density profile that satisfies both hydrostatic balance and our equation of state is reached.  Our simulated convection zones span from 0.72 R$_\odot$ to 0.965 R$_\odot$, extending to 25 Mm below the solar surface, and encompassing roughly 3.5 density scale heights.  Convective perturbations about this reference state remain small throughout the simulation, and thus we choose not to update the reference state, holding it fixed throughout each calculation.   

As the microscopic viscous and thermal diffusivities appropriate for plasma motions within the convection zone are beyond the capabilities of current computational models, our models should be viewed as large eddy simulations.   We adopt a radial profile for $\nu$ and $\kappa$ with each quantity diminishing in depth as $\nu\sim\rho^{-1/2}$.  Simulations are thus more diffusive near the surface where we expect the unresolved scales of convective motion to play a prominent role in the transport of heat and momentum. This formulation does not, however, capture any non-dissipative behavior of the unresolved scales.

Impenetrable and stress free boundary conditions are adopted for each simulation such that
\begin{equation}
v_r=\frac{\partial (v_{\theta}/r)}{\partial r}=\frac{\partial (v_{\phi}/r)}{\partial r}=0|_{r = r_{bot},r_{top}}.
\end{equation}
The radial entropy gradient is forced to vanish at the lower boundary of the convection zone, and the entropy perturbations are forced to vanish at the upper boundary, with
\begin{equation}
\frac{\partial S}{\partial r} = 0|_{r=r_{bot}},~~S = 0|_{r=r_{top}}.
\end{equation}
Thus, there is no diffusive entropy flux across the lower boundary.  Heat is transported across this boundary by the radiative heat flux instead, $\propto \kappa_r \del \overline{T}$, which drops to near zero at the upper boundary (see Fig.\ \ref{flux_bal} below).  The vanishing of the entropy gradient at the lower boundary implies that this is, by definition, the base of the CZ, with no overshoot region.  The entropy gradient steepens toward the top of the domain, where most of the convective driving occurs.  This will be discussed further in \S\ref{sec:heat}.  With the value of entropy pinned at the upper boundary, the entropy gradient there is free to vary with time.  As the simulation equilibrates, this gradient attains a statistically steady value that is sufficient to expel one solar luminosity through the upper boundary via thermal diffusion.  

A summary of our input parameter space for those cases rotating at the solar rate of $\Omega_\odot$=2.6$\times10^{-6}$ s$^{-1}$ is provided in Table \ref{inputs_table}. We vary the viscosities by a factor of four and the thermal diffusivities by a factor of two for these simulations.  The naming convention is such that the letter denotes the level of thermal diffusion, with \textit{A} corresponding to the highest thermal diffusivity of 3.2$\times$10$^{13}$ cm$^2$ s$^{-1}$ and \textit{C} corresponding to the lowest level (1.6$\times$10$^{13}$ cm s$^{-1}$).  The numbers \textit{0--3} indicate the level of viscous diffusion, with \textit{0} corresponding to the lowest kinematic viscosity (2$\times$10$^{12}$ cm s$^{-1}$), and \textit{3} corresponding to the highest value of 8$\times$10$^{12}$ cm s$^{-1}$.  Case B2, which lies in the center of our parameter space, serves as the point around which rotation rate is varied for the results presented in \S\ref{sec:regimes}.  We have also provided a summary of the resulting energy balances, as well as their relative ratios, for these cases in Table 2.  There, we report on the globally-averaged (over the full spherical shell) convective kinetic energy density (CKE), the energy associated with the differential rotation (DRKE), and the energy associated with the meridional circulation (MCKE). We define these as
\begin{equation}
	CKE =\frac{1}{2}\overline{\rho}\left[{(v_{r}-\langle{v_r}\rangle)^2 + (v_{\theta}-\langle{v_\theta}\rangle)^2 + (v_{\phi}-\langle{v_\phi}\rangle)^2 }\right],
\end{equation}
\begin{equation}
	MCKE =\frac{1}{2}\overline{\rho}\left[\langle{v_r}\rangle^2 + \langle{v_\theta}\rangle^2\right],
\end{equation}
and
\begin{equation}
	DRKE =\frac{1}{2}\overline{\rho}\left[\langle{v_\phi}\rangle^2\right],
\end{equation}
where the brackets denote averages over longitude.
%%%%%%%%%%%%%%%%%%%%%%%%%%%%%%%%%%%%%%%%%%%%%%%%%%%%%%%%%%%%%%%%
%%%%    Table 1
\begin{table}[t]
\label{inputs_table}\centering\small
\begin{tabular}[t]{lccccccc}\\
\multicolumn{8}{c}{TABLE 1}\\
\multicolumn{8}{c}{Variable Input Parameters}\\\hline\hline
 Case & $\nu_{top}$ & $\kappa_{top}$ & Pr & Ro& Ra$/10^{5}$& $\mathcal{R}^{*}$&$\ell_{max}$ \\\hline
A1 & 4 & 32 & 1/8  & 0.15& 1.79& 2.61& 340\\
A2 & 6 & 32 & 3/16 & 0.13& 1.19& 2.61& 170\\
A3 & 8 & 32 & 1/4  & 0.11& 0.89& 2.61& 85\\\hline
B0 & 2 & 24 & 1/12 & 0.16& 6.35& 3.48& 680\\
B0.5 & 3 & 24 & 1/8 & 0.16& 4.24& 3.48& 340\\
B1 & 4 & 24 & 1/6  & 0.15& 3.18& 3.48& 340\\
B2 & 6 & 24 & 1/4  & 0.13& 2.12& 3.48& 170\\
B3 & 8 & 24 & 1/3  & 0.12& 1.59& 3.48& 170\\\hline
C1 & 4 & 16 & 1/4  & 0.24& 7.14& 5.22& 340\\
C2 & 6 & 16 & 3/8  & 0.20& 4.76& 5.22& 340\\
C3 & 8 & 16 & 1/2  & 0.12& 3.57& 5.22& 170\\\hline
\end{tabular}
\tablecomments{ Input parameters for all cases.  Viscous and thermal diffusivities ($\nu$ and $\kappa$ respectively) are quoted in units of 10$^{12}$ cm$^2$ s$^{-1}$.  The Prandtl number, Pr, is constant with depth. The Rossby number Ro$=\tilde{v}/(2\Omega L)$ is quoted at mid-convection zone.  The Rayleigh number Ra$=(\partial\rho/\partial S) (\partial S/\partial r) g L^4 /(\rho\nu\kappa)$ is quoted at the top of the simulation. The length scale $L$ is taken to be the depth of the convection zone (1.72$\times$10$^{10}$ cm) in both cases. The modified Rayleigh Number $\mathcal{R}^{*} = g\left|dS_c/dr \right|/(c_p\Omega^2)$ is quoted at mid-convection zone.}
\end{table}

%%%%%%%%%%%%%%%%%%%%%%%%%%%%%%%%%%%%%%%%%%%%%5
%%%%    Table 2
\begin{table}[t]
\label{outputs_table}\centering\small
\begin{tabular}[t]{lccccc}\\
\multicolumn{6}{c}{TABLE 2}\\
\multicolumn{6}{c}{Energy Balance}\\\hline\hline
 Case  & CKE & DRKE & DRKE/CKE & MCKE & MCKE/CKE \\\hline
A1 & 2.09 & 2.94 & 1.4091 & 0.019 & 0.0091\\
A2 & 1.46 & 2.38 & 1.6241 & 0.012 & 0.0079\\
A3 & 1.22 & 1.77 & 1.4447 & 0.010 & 0.0080\\\hline
B0 & 4.39 & 0.59 & 0.1345 & 0.032 & 0.0073\\
B0.5 & 3.52 & 1.87 & 0.5305 & 0.026 & 0.0075\\
B1 & 2.99 & 1.73 & 0.5777 & 0.024 & 0.0079\\
B2 & 2.23 & 1.79 & 0.8024 & 0.017 & 0.0078\\
B3 & 1.96 & 1.31 & 0.6721 & 0.016 & 0.0084\\\hline
C1 & 4.66 & 9.92 & 2.1285 & 0.054 & 0.0115\\
C2 & 3.73 & 6.04 & 1.6202 & 0.044 & 0.0117\\
C3 & 2.90 & 0.18 & 0.0625 & 0.025 & 0.0085\\\hline
\end{tabular}
\tablecomments{ Globally averaged energy balance as realized in each run.  Energy densities have been averaged over 10 rotation periods at the end of each simulation, and are reported in units of 10$^6$ erg cm$^{-3}$}
\end{table}

Each model discussed in this paper has been evolved for several thousand days, with elapsed simulated time ranging from 7,000 days to 14,000 days depending on the individual case.  This corresponds to several hundred rotational and convective overturning times.  All simulations have been well equilibrated viscously and thermally. As Prandtl numbers are always less than unity, the viscous diffusion timescale is the most limiting of these two timescales.  With the exception of case B0, which has been run for one viscous diffusion time, all simulations have been run for at least two viscous diffusion times.  We find that case B0 is transitioning to an anti-solar differential rotation rate after one viscous diffusion time of evolution.  As we are able to explore anti-solar regimes adequately with the more computationally tractable cases C1 and C2, we have chosen not to pursue the evolution of B0 further, but include it in Tables 1 and 2, as well as Figure \ref{rossb_domega}, to help delineate the point of transition between solar and anti-solar behavior.

\section{Distinct Mean Flow Regimes for Fast and Slow Rotators}\label{sec:regimes}

\subsection{Identification of Mean Flow Regimes}\label{sec:idreg}

%%%%%%%%%%%%%%%%%%%%%%%%%%%%%%%%%%%%%%%%%%%%%%%%%%%%%%%%%%%%%%%%%%%%%
%  Figure 1:  differential rotation with changing rotation rate.
\begin{figure*}
\centerline{\epsfig{file=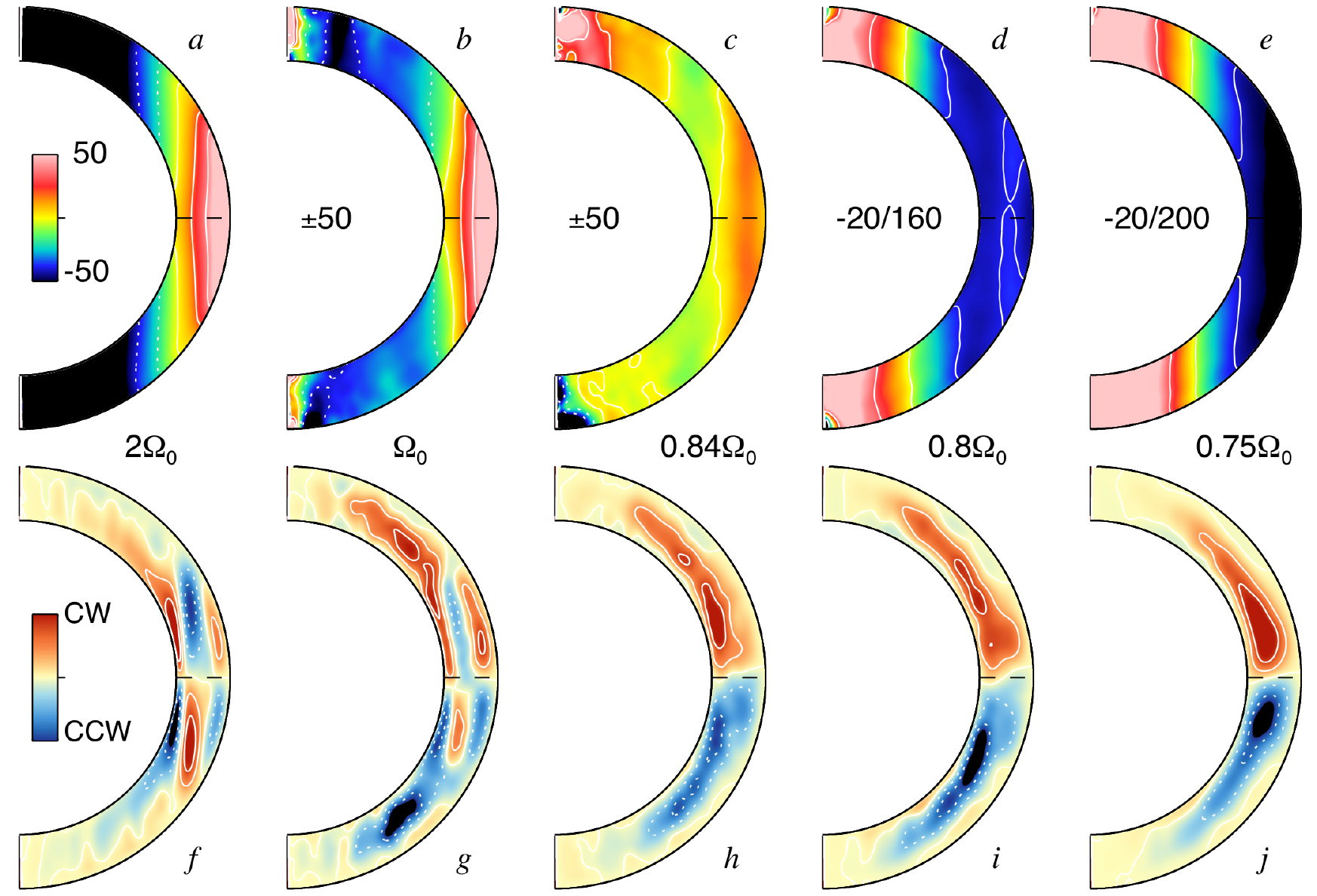,width=\textwidth}}
\caption{\label{spindown}(\textit{a}-\textit{e}) Differential rotation profiles ($\Omega-\Omega_{frame}$) for cases rotating at different fractions of the solar rotation rate $\Omega_0$ (indicated) achieved by spinning up and spinning down case B2, which rotates at the solar rate.  Rotation rate decreases from left to right, and  units are indicated in nHz. (\textit{f}-\textit{j}) Meridional circulation profiles corresponding to each of the differential rotation profiles in the upper row.  Red tones denote counter-clockwise motion, and blue tones, clockwise. Profiles have been averaged in time over roughly 200 days in each instance.  As the Sun is spun down, the polar regions develop prograde rotation, and the equatorial regions retrograde. Similarly, multi-cellular profiles of meridional circulation transition into single-celled profiles at low rotation rates.}
\end{figure*}

%%%%%%%%%%%%%%%%%%%%%%%%%%%%%%%%%%%%%%%%%%%%%%%%%%%%%%%%%%%%%%%%%%%%%%%%%%%

We find that the structure of meridional circulation realized in our convection zone models is intimately linked to the differential rotation established.  This relationship is most readily illustrated through a series of experiments wherein the stellar rotation rate is varied.  
Using Case B2 as our starting point, we have explored rotation rates ranging from 2$\Omega_\odot$ to 0.75$\Omega_\odot$.

We plot the resulting differential rotation (\textit{upper row}) and meridional circulation (\textit{lower row}) profiles for this series of simulations in Fig. \ref{spindown}  At the higher rotation rates, most notably 2$\Omega_\odot$ and $\Omega_\odot$, the resulting differential rotation is decidedly solar-like, possessing a fast equator and retrograde poles (see \S\ref{sec:models}).  We note that contours in these simulations are noticeably cylindrical with respect to the Sun.  
This may be attributed to the lack of an underlying stable zone which is thought to promote conical $\Omega$ profiles by inducing baroclinic torques \citep{rempe05,balbu12}.
At rotation rates less than $\Omega_\odot$, the solar-like differential rotation transitions to a state characterized by a weakly retrograde equator and rapidly prograde poles.  These prograde polar states tend to possess a much stronger equator-to-pole rotational contrast than their solar-like counterparts. Similar results were also reported by other authors, including \cite{gasti13,gasti14}, \cite{guerr13}, \cite{kapyl14}.

As the lower row of Fig. \ref{spindown} is traversed from left to right, circulatory patterns assume two distinct morphologies and exhibit a clear correlation with the structure of the differential rotation.  In the fast-equator cases, the low latitudes are characterized by a counter cell of circulation occurring near mid-convection zone.  High latitudes in these cases are largely characterized by a single cell of circulation (poleward at the surface), with some hints of multi-cellular behavior at high latitudes for the case with the fastest rotation.  As the rotation rate of the system decreases, the high latitude cell extends to lower latitudes, ultimately supplanting the counter cell in the equatorial regions for rotation rates lower than $\Omega_\odot$.   The shift from multi-cellular MC profiles in the solar regime to single-celled profiles in the anti-solar regime has also been noted by other authors \citep{guerr13,gasti14,kapyl14}.

%%%%%%%%%%%%%%%%%%%%%%%%%%%%%%%%%%%%%%%%%%%%%%%%%%%%%%%%%%%%%%
\begin{figure*}
\centerline{\epsfig{file=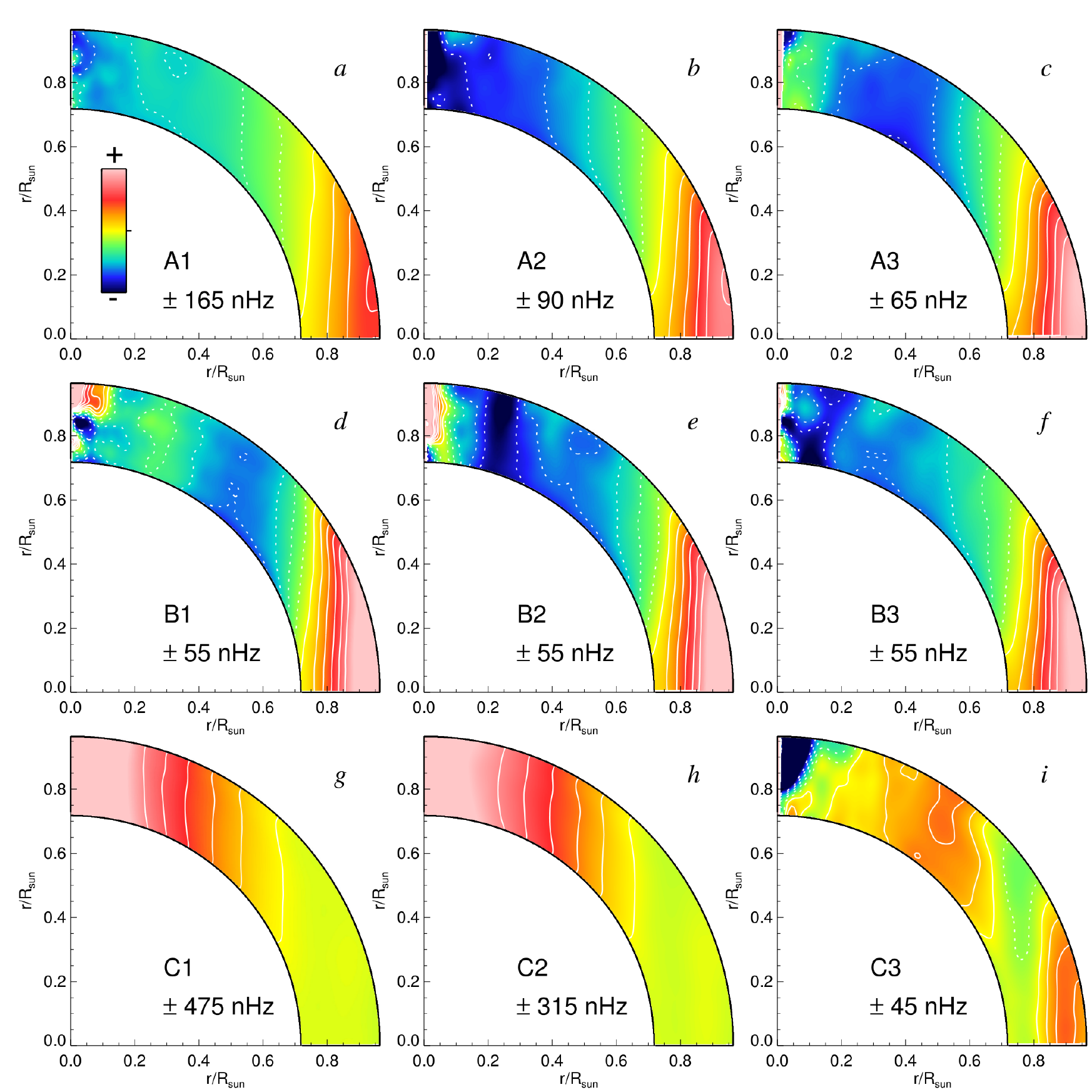,width=\textwidth}}
\caption{\label{dr_grid} Differential rotation profiles ($\Omega-\Omega_\odot$) for the \textit{a},\textit{b}, and \textit{c} series of simulations.  Case names are indicated and are arranged such that viscosity $\nu$ increases from left to right, and thermal diffusivity $\kappa$ increases from bottom to top.  The most laminar case (A3) appears in the upper right, and the most turbulent case (C1) appears in the lower left.  Profiles have been averaged in time over roughly 200 days in each instance.  A clear transition between solar-like and anti-solar rotation occurs between the middle and bottom rows (i.e. between the $B$ and $C$ series cases.)}
\end{figure*}

%%%%%%%%%%%%%%%%%%%%%%%%%%%%%%%%%%%%%%%%%%%%%%%%%%%%%%%%%%%%%%

\subsection{Transition Between Mean Flow Regimes at Fixed $\Omega_0$}

The series of simulations illustrated in Fig.\ \ref{spindown} possess a common mid-core rms velocity amplitude of about 100 m s$^{-1}$.  The transition in mean-flow regimes spanned by those simulations is thus most naturally characterized by a transition in timescale ratios, namely the ratio of the convective overturning time to the rotation period.  This is quantified by the Rossby number, introduced in \S\ref{sec:intro}.  We can similarly span this transition by holding $\Omega$ fixed and varying the amplitude of the convective velocities.   For systems with a fixed luminosity, such as those under consideration here, this is most naturally accomplished by varying the transport coefficients $\nu$ and $\kappa$, and through them the Rayleigh number $Ra$.  In so doing, we can also explore how our transport parameters $\nu$ and $\kappa$ contribute to the balances realized in the slowly- and rapidly-rotating regimes by sampling a range of resulting Reynolds numbers.

Others have shown that the transition between the solar and anti-solar rotation regimes can be achieved by fixing the rotation rate and changing the convective driving by varying the Rayleigh number \citep{gilma77,aurno07,guerr13,gasti14,kapyl14}.  The key parameter in determining the transition is the Rossby number.  However, since this is an output of the simulation rather than an input, the transition is often described in terms of the modified Rayleigh number $\mathcal{R}^*$ (see Table \ref{inputs_table}), which is a measure of the relative strengths the buoyancy and Coriolis force \citep{gasti13}.  The simulations presented in this section will be used in sections \ref{sec:interp} to explore the structure and origin of the meridional circulation, a topic that has received less attention in previous studies.  While our demonstration of the solar/anti-solar transition is itself not new, we find it useful to revisit some aspects of this transition as they also bear on the resulting meridional circulation.

%%%%%%%%%%%%%%%%%%%%%%%%%%%%%%%%%%%%%%%%%%%%%%%%%%%%%%%%%%%%%%
%   Figure X:  Rossby Number
\begin{figure}
\centerline{\epsfig{file=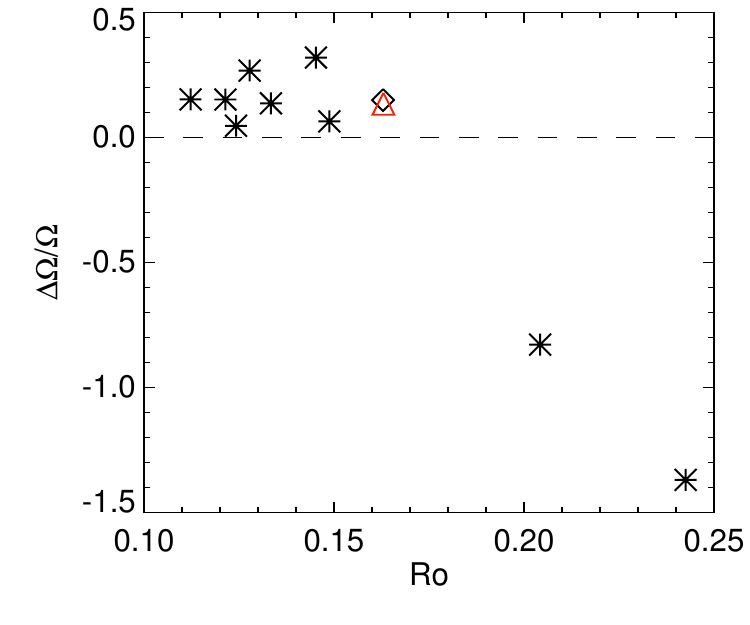,width=\linewidth}}
\caption{\label{rossb_domega}  Pole-to-equator differential rotation contrast $\Delta\Omega/\Omega$ as a function of Rossby number $Ro$ for the $A-C$ series of simulations. Transitional case B0.5 is indicated with a hollow diamond.  The slightly more turbulent case B0, which is transitioning to an anti-solar state, is indicated by the red triangle.  $Ro$ was computed at mid-convection zone in all cases.  Equatorial and polar rotation rates were calculated by averaging zonal velocity in both longitude and depth, and then over a latitude range of 0$^\circ$N-20$^\circ$N latitude and 70$^\circ$N-90$^\circ$N respectively.  Anti-solar cases tend to arise when $Ro$ is greater than about 0.163.   }
\end{figure}

%%%%%%%%%%%%%%%%%%%%%%%%%%%%%%%%%%%%%%%%%%%%%%%%%%%%%%%%%%%%%%

%%%%%%%%%%%%%%%%%%%%%%%%%%%%%%%%%%%%%%%%%%%%%%%%%%%%%%%%%%%%%%
\begin{figure*}
\centerline{\epsfig{file=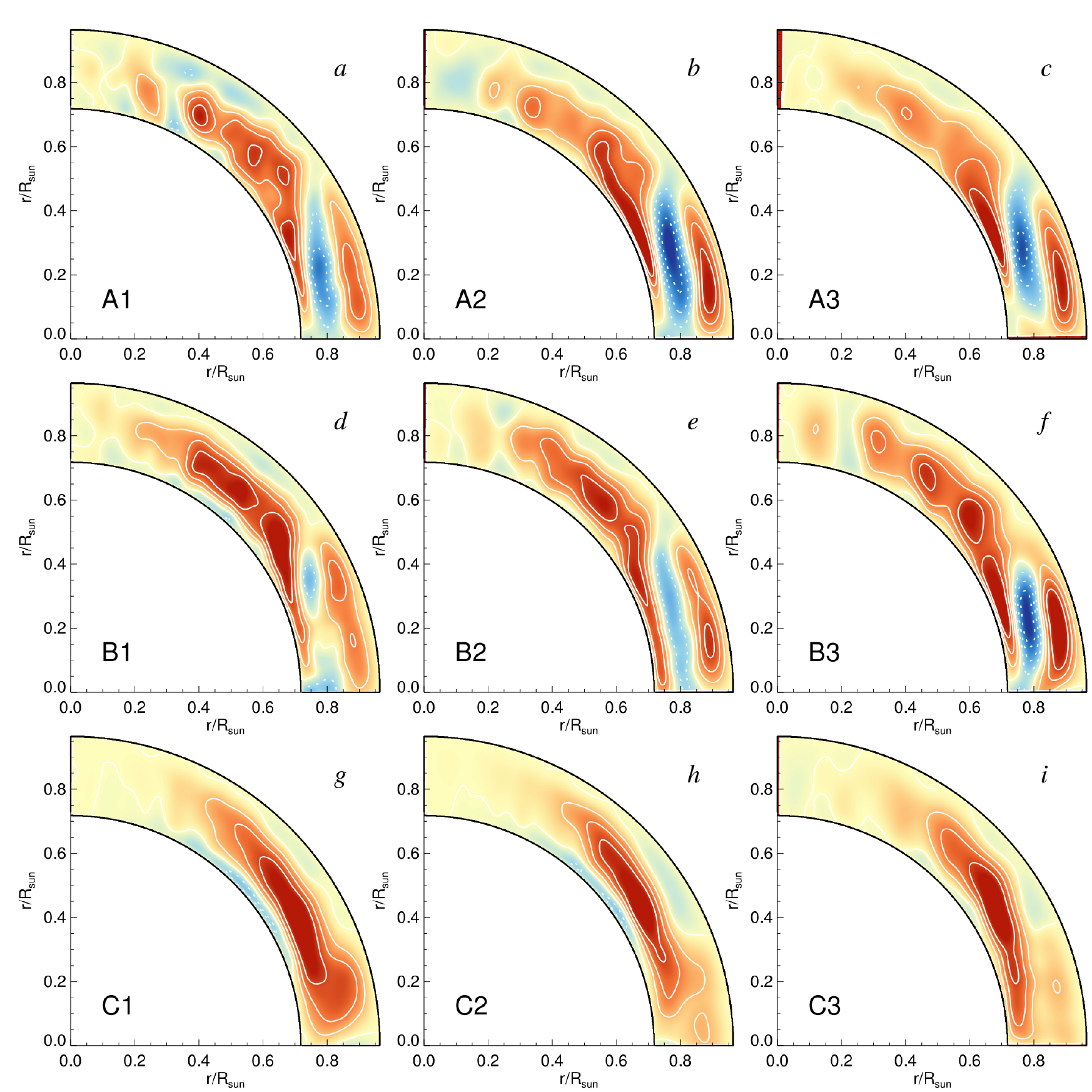,width=\textwidth}}
\caption{\label{circ_grid} Contours of meridional circulation streamlines, with streamfunction underlay, for cases in the \textit{A}-\textit{C} series as indicated within each panel.  The layout and time averaging are as in Fig. \ref{dr_grid}.  Red tones denote counter-clockwise flow, and blue tones clockwise.  As with cases where the rotation rate is varied, multi-cellular circulations arise whenever solar-like differential rotation profiles are present.}
\end{figure*}

\begin{figure*}
\centerline{\epsfig{file=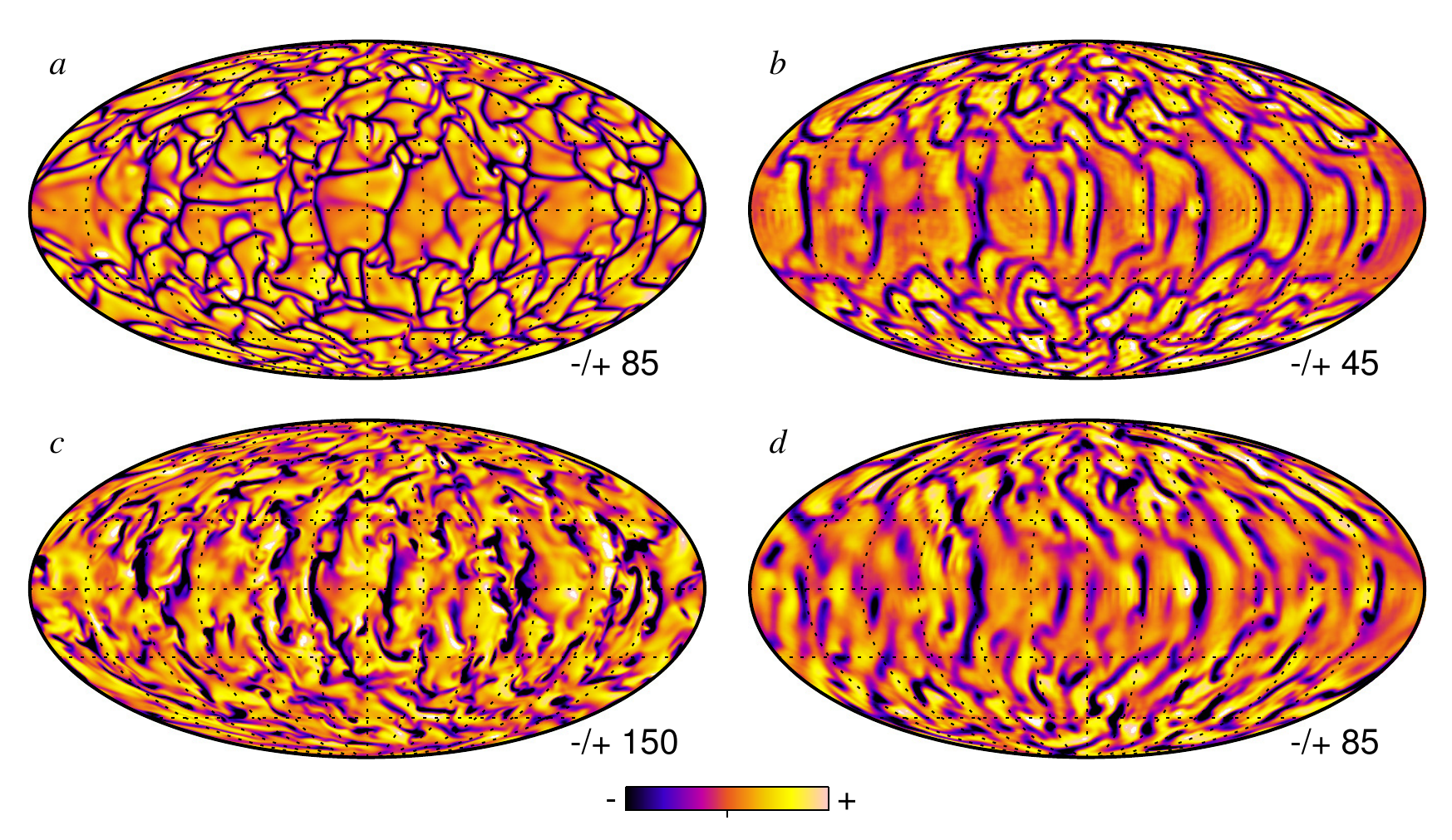,width=\textwidth}}
\caption{\label{solaranti} Radial velocity at one instant in time following equilibration for cases C1 (left column) and A3 (right column). The upper and lower rows correspond to radial velocity near the surface and in the mid CZ respectively.  Convective cells in the solar-like case A3 exhibit a prograde tilt in the $\phi$-direction.  Those in anti-solar case C1 are tilted in the opposite sense.  The limits of the color table normalization are indicated adjacent to each image, with units provided in m s$^{-1}$.}
\end{figure*}
%%%%%%%%%%%%%%%%%%%%%%%%%%%%%%%%%%%%%%%%%%%%%%%%%%%%%%%%%%%%%%

It is worth mentioning that the Rayleigh number in systems such as these, where entropy is held fixed at one boundary, and the entropy gradient fixed at the other, possesses an implicit dependence on $\kappa$ not found in classical Rayleigh-B\'{e}nard convection (RBC).  The entropy contrast in models such as ours, analogous to the temperature contrast in RBC systems, is not held fixed, but is an output quantity, much as the Reynolds number.  This entropy contrast is built entirely in the upper boundary layer, resulting from the entropy gradient realized there and the width of that boundary layer.  Convective driving is thus accomplished primarily in the upper boundary layers of these systems, a situation more analogous to the Sun than RBC flow where driving occurs at the upper and lower boundaries.

While entropy is fixed at the top, the entropy gradient is implicitly determined by the solar luminosity, being forced to satisfy, in a time-averaged sense, the relation 
\begin{equation}
\label{eq:entropy_gradient}
\avg{\rho}\avg{T}\kappa\frac{\partial s}{\partial r}\vert_{r=r_{top}} = \frac{L_\odot}{4\pi r^2_{top}}.
\end{equation}
We thus choose to define the our Rayleigh numbers using $\partial{s}/\partial{r}$ at the upper boundary because this quantity is a well-defined simulation input.  The consequence is that our Rayleigh number now varies as $\kappa^{-2}$ instead of the more traditional $\kappa^{-1}$.

The variation of differential rotation for each case is illustrated in Fig. \ref{dr_grid}. There, profiles of angular velocity, averaged over several overturning times, are laid out in a grid, with $\kappa$ increasing along the vertical axis and $\nu$ increasing along the horizontal axis.  The upper row is comprised by simulations from Series A, and the lower row by Series C simulations.  Convection models from Series A and B each exhibit solar-like differential rotation, whereas series C tends to exhibit anti-solar rotation, with case C3 (lower right) possibly representing a transition point between the fast equator and fast pole regimes.   Variation of the differential rotation with $\nu$ is present, but the effect is greatly reduced with respect to that of $\kappa$.  This can be attributed to the $\nu^{-1} \kappa^{-2}$ scaling of the Rayleigh number noted above.

As noted above, the key parameter in determining the transition between mean flow regimes is the Rossby number $Ro$.  Fig. \ref{rossb_domega} shows the relative differential rotation, $\Delta \Omega/\Omega_0 =(\Omega_{eq}-\Omega_{pole})/\Omega_0$, in all simulations versus $Ro$, exhibiting a clear transition at $R_o \sim$ 0.17.  This transitional value is lower than the $R_o \sim 1$ value quoted in previous papers \citep[e.g.][]{gasti13,gasti14}.  We attribute this discrepancy to the ambiguity in defining $Ro$.  Here we use the depth of the CZ as the length scale $L$ and the rms value of the non-axisymmetric velocity in the mid-CZ $v^\prime_{rms}$ as the velocity scale $U$.  By contrast, \citep{gasti14} use the volume-averaged value of $v^\prime_{rms}$ for $U$ and a value of $L$ based on the peak of the kinetic energy spectrum.  This tends to shift $U$ higher and $L$ lower, resulting in a larger value of $Ro$.  We refer the reader to \citep{gasti14} for a discussion of how the transitional value of $Ro$ depends on precisely how $Ro$ is defined.

The meridional circulation profiles corresponding to the simulations shown in Fig.\ \ref{dr_grid} are shown in Fig.\ \ref{circ_grid}.  They exhibit a correlation with the differential rotation similar to that discussed in \S\ref{sec:idreg}.  Single-celled profiles of circulation appear in all of the C-series cases (i.e. those systems with fast poles), while those of the A-B series are multi-cellular in the equatorial regions.

Convection in the solar and anti-solar regimes exhibits noticeably different morphologies as well.  Fig. \ref{solaranti} illustrates the radial velocities in the upper and mid-convection zones of case C1 (\textit{a,c}) and case A3 (\textit{b,d}).  Case C1, which possess a anti-solar-like differential rotation, also exhibits velocity amplitudes that are roughly twice that of A3. Elongated convective cells in the equatorial regions tend to tilt in the negative $\phi$-direction for case C1.  In addition to lower velocity amplitudes, 
Case A3 possesses much more prominent columnar structuring of its convection, with a tilting evident in the positive $\phi$-direction.  

An additional effect is evident when comparing the flows of case A3 to those of case C1, namely that the convection exhibits a much broader range of spatial scales.  Case C3 is thus both more turbulent and more strongly driven.  

\subsection{A Transitional Regime}\label{sec:transition}
One might ask if the transition in mean-flow regimes is purely a result of rotational constraint, or if the level of turbulence might also play some role.  In particular, by admitting much smaller spatial scales into the simulation, are the correlations of $v_r$ and $v_\phi$ necessary to establish a fast equator being disrupted? In other words, might the Reynolds number $R_e = U L / \nu$ play a role in addition to the Rayleigh number? In order to explore these questions we have extended the simulations of Fig. \ref{dr_grid} to include the case B0.5.  This case possesses a $\kappa$ identical to the other B-series cases, but a $\nu$ that is lower than any of of the C-series.  Case B0.5 was initiated by lowering the diffusivity in the mature case B1, which had itself been evolved for a total of two viscous diffusion times.

An overview of the flows in Case B0.5 is shown in Fig. \ref{turb_solar}.  The convective patterns are small-scale in nature, like C3, but do not exhibit the retrograde tilting characteristic of the patterns in this case.  Case B0.5 possesses a decidedly solar-like differential rotation, but, surprisingly, seems to exhibit a single circulation cell within each hemisphere.  Moreover, the flow amplitudes indicated near the surface are similar to those of case C3, as are the mid-convection zone velocities (not shown).  Case B0.5 does possess a somewhat lower Rayleigh number, and a different Prandtl number than case C3, however, and so the differences between these two cases are likely to be subtle ones.  It would seem that case B0.5 is just straddling the transition between the solar-like and anti-solar regimes of differential rotation. 

\cite{gasti14} and \cite{kapyl14} have recently shown that the behavior of the differential rotation
near the solar/anti-solar transition can depend on the history of the simulation, representing a type of hysteresis.  For a given transitional $Ro$, the simulation may exhibit either solar or anti-solar DR.  Though we have observed unusual mean flows near the transition with case B0.5, we have not searched explicitly for this bistability and hysteresis.  It is possible that case B0.5 itself is slowly undergoing a transition from solar-like to anti-solar differential rotation.  However, this case has been run for 18,000 days, about a factor of four longer than the time scale for viscous diffusion across the convection zone, and we see no signs yet that it is undergoing such a transition.  The globally-averaged energy densities associated with the differential rotation (DRKE) and meridional circulation (MCKE) for case B0.5 are shown in Fig. \ref{energy_evol}. The mean flows have varied by as much as 30\% over this interval, but with no apparent long-term trends.    We will discuss the significance of case B0.5 further in \S\ref{sec:sas}.

%%%%%%%%%%%%%%%%%%%%%%%%%%%%%%%%%%%%%%%%%%%%%%%%%%%%%%%%%%%
%%% Figure 6 (Transitional Case Snapshot) - 1800 day average for Mean Flows

\begin{figure}
\centerline{\epsfig{file=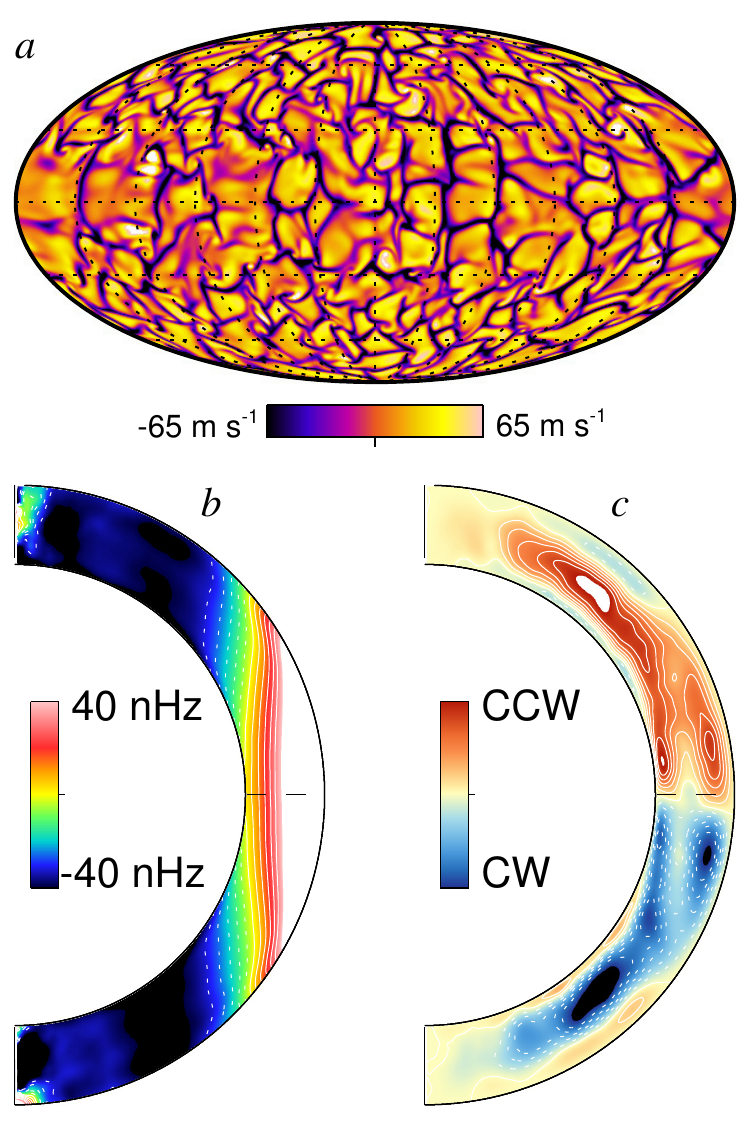,width=\linewidth}}
\caption{\label{turb_solar}Sampling of flows for transitional case B0.5.  (\textit{a})  Radial velocity snapshot near the upper boundary,  Upflows are indicated in yellow, downflows in blue.  (\textit{b}) Accompanying differential rotation ($\Omega-\Omega_\odot$), time-averaged over 1800 days at the end of the simulation, with prograde rotation shown in red, and retrograde rotation in blue.  (\textit{c}) Meridional circulation streamlines for the same time period.  Clockwise flow is indicated by blue underlay, and counter-clock wise flow by red underlay.  This case demonstrates a prograde rotating equator in the absence of multi-cellular meridional circulations within each hemisphere.} 
\end{figure}

%%%%%%%%%%%%%%%%%%%%%%%%%%%%%%%%%%%%%%%%%%%%%%%%%%%%%%%%%%%%%%%
%%% Figure New 7 (transitional case energy evolution)
\begin{figure*}[t]
\centerline{\epsfig{file=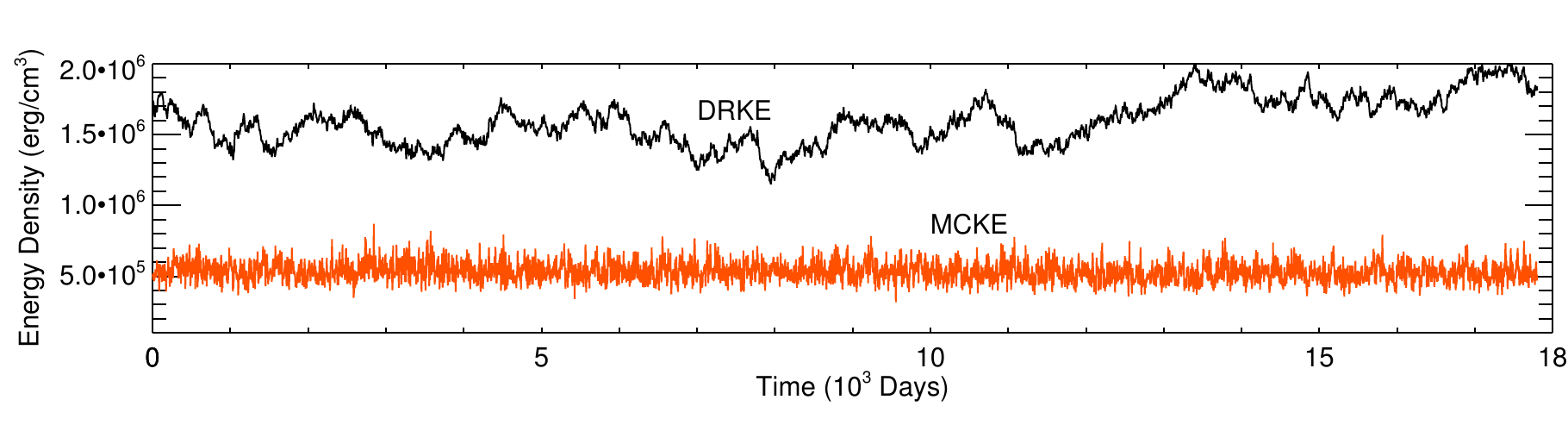,width=\linewidth}}
\caption{\label{energy_evol}Evolution of mean-flow energy densities for transitional case B0.5.  The energy in the differential rotation (DRKE) is depicted in black, and that of the meridional circulation (MCKE) in red.  DRKE exhibits large (roughly 30\%) variations with respect to the mean over this interval, but no long-term trends are evident.} 
\end{figure*}

%MMMMMMMMMMMMMMMMMMMMMMMMMMMMMMMMMMMMMMMMMMMMMMMMMMMMMMMMMMMMMMMMMMMMMM
\section{Maintenance of Mean Flows}\label{sec:interp}

In this section we argue that the distinct mean flow regimes described in 
\S\ref{sec:regimes} arise from a change in the nature of the convective
angular momentum transport.  Furthermore, we argue that it is the
convective angular momentum transport that largely determines the
mean meridional flow profiles in simulations as well as stars.

\subsection{Dynamical Balances and Gyroscopic Pumping}\label{sec:gp}

In an anelastic (low Mach number) system, the mass flux $\rh \vv$ is divergenceless
and the time evolution of the mean meridional flow $\rh \left<\vv_m\right>$ is 
governed by the zonal vorticity equation, which may be written as
\citep[][hereafter MH11]{miesc11}  
\begin{equation}\label{eq:vort}
\frac{\pd }{\pd t} \left<\omega_\phi\right> = \lambda \frac{\pd \Omega^2}{\pd z}
- \frac{g}{r C_P} \frac{\pd \left<S\right>}{\pd \theta} + \ldots ~~~,
\end{equation}
where $\omega_\phi = (\curl \vv_m) \bdot \uvp$ is the zonal component of 
the mean vorticity, $\lambda = r\sin\theta$ and $z = r \cos\theta$ are 
the radial and axial coordinates in a (rotating) cylindrical coordinate 
system, and $\Omega = \Omega_0 + \lambda^{-1} \left<v_\phi\right>$ is the total
angular velocity, including uniform and differential rotation components.  We have adopted the convention that a subscript $m$ on a vector denotes the radial {\mbox{\boldmath $\hat{r}$}} and latitudinal {\mbox{\boldmath $\hat{\theta}$}} components of a vector.
We have focused here on the two dominant contributors to the meridional
forcing, namely the inertia of the differential rotation and the baroclinicity
of the mean stratification, which make up the first and second terms on 
the right-hand-side of (\ref{eq:vort}) respectively.  These are sufficient
to illustrate the essential dynamics.  We have neglected the meridional components of the Reynolds stress ($\left<\avg{\rho}\vv_m'\vv_m'\right>$; primes denote fluctuations about the longitudinal mean), the baroclinicity of 
thermal fluctuations, the viscous stress, and quadratic terms in $\vv_m$.
All of these are thought to be negligible in simulations and stars.
We have also neglected the Lorentz force, which is less justified
in stars but appropriate here since the simulations we consider
are non-magnetic.  Full expressions are given in MH11 
\cite[see also][]{kitch95,rempe05}.

In a steady state eq.\ (\ref{eq:vort}) gives the familiar 
thermal wind balance (TWB) equation \citep{kitch95,ellio00,robin01,brun02,rempe05,miesc06,balbu09,brun10}
\begin{equation}\label{eq:twb}
\frac{\pd \Omega^2}{\pd z} = \frac{g}{\lambda r C_P} \frac{\pd \left<S\right>}{\pd \theta} ~~~.
\end{equation}
Here the two dominant meridional forcing terms balance, and latitudinal thermal gradients can sustain
non-cylindrical rotation profiles as in the Sun ($\pd \Omega / \pd z \neq 0$).  

It is important to note that in a steady state the meridional flow itself drops
out of equation (\ref{eq:twb}).  Thus, it cannot be used to determine the 
MC profile even if other quantities such as $\Omega$ and $S$ are known.
This ceases to be the case if other terms are included in the balance,
most notably the meridional components of the Reynolds stress.  In mean 
field theory, this term is often represented as a turbulent diffusion.
In short, inclusion of a turbulent diffusion term breaks the degeneracy
of eq.\ (\ref{eq:twb}) with respect to the meridional flow and can 
largely determine the MC profile that is achieved in many mean-field
models \citep{kuker11,kitch12,dikpa14}.  However, such solutions are
sensitive to the nature of the imposed parameterizations, such as the
depth dependence of the turbulent viscosity and departures from TWB
in the upper and lower boundary layers, where viscous stresses can 
ultimately determine the global structure and amplitude of the flow.

There is an alternative way to break the degeneracy of eq.\ (\ref{eq:twb})
that provides a better explanation for how the MC is established in
our 3-D, time-dependent convection simulations.  The mechanism is 
known as gyroscopic pumping and can be illustrated by 
considering the zonal force balance which can be expressed as
(MH11)
\begin{equation}\label{eq:gp}
\rh \left<\vv_m\right> \bdot \del {\cal L} = {\cal F}  
\approx - \dv {\bf F}_{RS} ~~~,
\end{equation}
where ${\cal L} = \lambda^2 \Omega$ is the specific angular momentum,
${\cal F}$ is the net axial torque given by 
\begin{equation}\label{eq:torque}
{\cal F} = - \dv \left[\rh \lambda \left<\vv_m^\prime v_\phi^\prime\right> 
- \rh \nu \lambda^2 \del \Omega\right]  
= - \dv \left[{\bf F}_{RS} + {\bf F}_{VD}\right]
~~~,
\end{equation}
and ${\bf F}_{RS}$ is the convective transport of angular momentum, given by
\begin{equation}\label{eq:frs_nick}
{\bf F}_{RS} = \rh \lambda \left<\vv_m^\prime v_\phi^\prime\right>~~~.
\end{equation}
In stars and in recent high-resolution convection simulations such as 
those presented here, the Reynolds stress component ${\bf F}_{RS}$
generally dominates over the viscous component ${\bf F}_{VD}$.
The dynamical balances (\ref{eq:twb}) and (\ref{eq:gp}) should be regarded
as temporal averages as well as longitudinal averages, concerning the
persistent mean flows that exist amid stochastic fluctuations.

%%%%%%%%%%%%%%%%%%%%%%%

In convection simulations and in helioseismic inversions, $\del {\cal L}$ is 
oriented away from the rotation axis, parallel to the cylindrical radius
$\del \lambda$ (MH11).  Thus, a convergence of the convective angular 
momentum flux (${\cal F} > 0$) will induce a meridional flow
$\left<\vv_m\right>$ directed away from the rotation axis and a 
divergence (${\cal F} < 0$) will induce a flow toward the rotation
axis.  This is the phenomenon of gyroscopic pumping as discussed 
in a solar context by MH11 
\citep[see also][]{hayne91,mcint98,mcint07,garau09,garau10,miesc12b}.  

Gyroscopic pumping is mediated by the axial component of the rotational
shear $\pd \Omega / \pd z$, but it can be sustained even if the 
steady-state rotation profile is strictly cylindrical; for an analytic 
illustration, see Appx.\ B of MH11.  Indeed, this is also demonstrated
in the simulations presented here; many exhibit $\Omega$ profiles
that are nearly cylindrical yet sustain persistent, well-established
meridional flow profiles (e.g.\ Fig.\ \ref{spindown}).  Intuitively,
one can attribute this to the efficiency and robustness of Taylor-Proudman
balance; in the barotropic limit ($\pd \left<S\right>/\pd \theta = 0$)
any axial variation of the net torque ($\pd {\cal F}/\pd z \neq 0$)
will induce a transient axial shear ($\pd \Omega / \pd z \neq 0$)
that will almost immediately\footnote{on a timescale 
$\tau \sim (2\Omega)^{-1} \sim P_{rot}/(4\pi)$, where $P_{rot}$ 
is the rotation period. For the Sun this is about 2.2 days.} be wiped 
out by an induced meridional flow, via equation (\ref{eq:vort}).  In 
this way, a steady meridional flow can be maintained that continually 
replenishes angular momentum extracted or imparted by the net torque 
${\cal F}$ by advecting angular momentum across local ${\cal L}$ 
isosurfaces as expressed by eq.\ (\ref{eq:gp}).

Eq. (\ref{eq:gp}) predicts a direct link between the meridional flow profile and the angular
momentum transport by the convective Reynolds stress.
This link is strongest for the low-Rossby number limit in which 
the uniform rotation component of $\Omega$ dominates ${\cal L}$.
In this limit we have
\begin{equation}
\Psi(\lambda,z) = \frac{1}{2 \lambda \Omega_0} ~ 
\int_{z_b}^z {\cal F}(\lambda,z^\prime) ~ dz^\prime ~~~,
\end{equation}
where $z_b = (R^2 - \lambda^2)^{1/2}$ (MH11) and $\Psi$ is the 
cylindrical mass flux streamfunction, defined by
\begin{equation}\label{eq:vlambda}
\left< \avg{\rho} v_\lambda\right> = \frac{\partial \Psi}{\partial z}
\mbox{\hspace{.1in}},
\left< \avg{\rho} v_z\right> = - \frac{1}{\lambda} \frac{\partial}{\partial \lambda} \left(\lambda \Psi\right)
~~~.
\end{equation}
Recall that $\lambda$ is the cylindrical radius ($\lambda = r \sin\theta$).  Note that $\Psi$ is related to the 
mass flux streamfunction $W$ as follows
\begin{equation}
\Psi = \frac{1}{r} \frac{\pd \left<W\right>}{\pd \theta} ~~~.
\end{equation}

In summary, there are two competing mechanisms for breaking the degeneracy 
of (\ref{eq:twb}) and establishing the steady-state MC; we will refer to 
them as the meridional Reynolds stress (RS) and gyroscopic pumping.  
Both rely on the RS tensor ${\cal R}_{i j} = \rh \left<v^\prime_i v^\prime_j\right>$.
However, the first mechanism relies on the meridional components of ${\cal R}$
(${\cal R}_{r r}$, ${\cal R}_{\theta r}$, ${\cal R}_{\phi r}$,  
 ${\cal R}_{r \theta}$, ${\cal R}_{\theta \theta}$, ${\cal R}_{\phi \theta}$),
whereas the second relies on the zonal components 
(${\cal R}_{r \phi}$, ${\cal R}_{\theta \phi}$, ${\cal R}_{\phi \phi}$).
In the next section we will demonstrate that it is this latter mechanism,
namely gyroscopic pumping, that determines the MC profiles in our
convection simulations.

However, before proceeding, we first elaborate on these two mechanisms
within the context of mean-field models.  In our convection simulations
we capture all of the RS components explicitly so both of these mechanisms
occur naturally.  Both mechanisms may also be captured in mean-field
models by parameterizing the RS based on convection simulations as
presented here or on alternative phenomenological arguments or 
turbulence models.  As mentioned above, the meridional RS 
mechanism may be captured by solving the steady-state zonal vorticity 
equation \citep{kuker11,kitch12,dikpa14}.  However, in order to capture the 
gyroscopic pumping mechanism it is essential to include the zonal momentum 
equation and to follow the time-dependent approach to the steady state 
\cite{rempe06}.  

Though both mechanisms can and have been realized in mean-field models, 
the gyroscopic pumping mechanism is less sensitive to imposed
mean-field parameterizations.  There are two reasons for this.
First, the meridional MC mechanism relies on second-order departures 
from the primary meridional force balance, which in the bulk of the
convection zone is TWB, eq.\ (\ref{eq:twb}).  By contrast, the zonal
RS enters into the lowest-order zonal force balance, eq.\ (\ref{eq:gp}).
The second reason is that the meridional MC mechanism is sensitive
to the functional form of the RS parameterization, ${\cal R}(\Psi,\Omega)$.
Thus, implementing results from a convection simulation would require
parameterizing the computed RS tensor.  However, in the gyroscopic
pumping mechanism one must only know the axial torque ${\cal F}$.
This may in principle be decoupled from the mean flows without
the need for any explicit parameterizations.  The zonal 
components of the RS do likely depend on the amplitude of the 
differential rotation, however, and a realistic mean-field model should take
this into account.

%>>>>>>>>>>>>>>>>>>>>>>>>>>>>>>>>>>>>>>>>>>>>>>>>>>>>>>>>>>>>>>

\subsection{Convective Angular Momentum Transport}\label{sec:amom}

In \S\ref{sec:gp} we argued that the meridional flow profiles in our
simulations are largely maintained by the convective angular momentum
transport ${\bf F}_{RS}$ through the mechanism of gyroscopic pumping.  In this section we 
assess this argument and its implications by exploring the detailed 
nature of ${\bf F}_{RS}$ and how it relates to the meridional flow
profiles we find in the simulations.

\begin{figure*}
\centerline{\epsfig{file=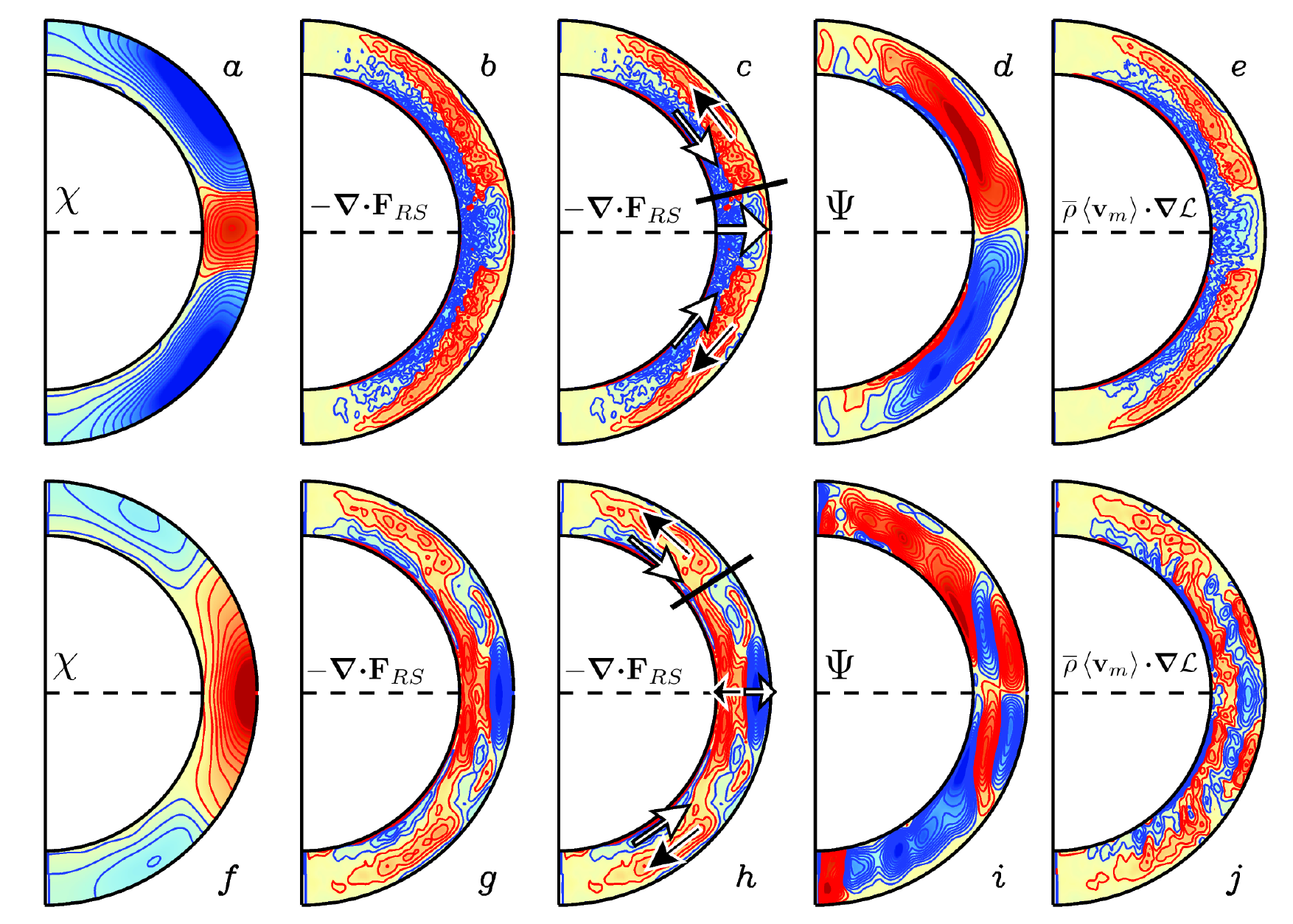,width=\textwidth}}
\caption{\label{fig:chi1} Maintenance of meridional flow. Illustrative
examples are shown for the rapidly-rotating regime (upper row, case C1)
and the slowly-rotating regime (lower row, case A3).  (\textit{a}, \textit{f})
Transport potential $\chi$, increasing from -$10^{26}$ g cm s$^{-2}$ (blue)
to $10^{26}$ g cm s$^{-2}$ (red).
(\textit{b}, \textit{g}) Divergence of the convective angular momentum transport,
ranging from -$10^7$ g cm$^{-1}$ s$^{-2}$ (blue) to $10^7$ g cm$^{-1}$ s$^{-2}$ (red).
(\textit{c}, \textit{h}) same as frames (\textit{b}, \textit{g}) but
with arrows indicating the direction of the induced meridional flows.
Slanted lines in the northern hemisphere delineate high and low-latitude regimes.
(\textit{d}, \textit{i}) Meridional flow shown as streamlines of the
mass flux $\Psi$. Red denotes clockwise flow ($\Psi > 0$) and blue denotes 
counter-clockwise flow ($\Psi < 0$).  Values of $\Psi$ range from 
$\pm 4\times 10^{11}$ g cm$^{-1}$ s$^{-1}$ in (\textit{d}) and 
$\pm 9\times 10^{10}$ g cm$^{-1}$ s$^{-1}$ in (\textit{i}).
(\textit{e}, \textit{j}) Angular momentum transport by the 
meridional flow which approximately balances the Reynolds stress divergence in 
frames (\textit{b}, \textit{g}), as expressed in eq.\ (\ref{eq:gp}).  Color 
table as in frames (\textit{b}, \textit{g}).}

\end{figure*}

In our discussion, we find it instructive to introduce a Helmholtz decomposition
of the convective angular momentum transport ${\bf F}_{RS}$
as follows:
\begin{equation}\label{eq:helm}
{\bf F}_{RS} = \del \chi + \curl \left(\Lambda \uvp\right)  ~~~.
\end{equation}
Such a decomposition is valid for any arbitrary
two-dimensional (axisymmetric) vector.  Its utility
is apparent when we see that only the first term contributes
to the gyroscopic pumping equation (\ref{eq:gp}).
More generally, only the divergent component of the angular 
momentum flux can contribute to the time evolution of 
$\Omega$, and thus the mean flow profiles that are 
ultimately achieved.  One can derive the transport
potential $\chi$ from ${\bf F}_{RS}$ by taking the
divergence of (\ref{eq:helm}) and solving a Poisson 
equation:
\begin{equation}\label{eq:poisson}
\nabla^2 \chi = \dv {\bf F}_{RS} ~~~.
\end{equation}

Figure \ref{fig:chi1} demonstrates the nature of the convective
angular momentum transport in two representative cases, chosen
to illustrate the low and high Rossby number regimes (upper and 
lower rows respectively).  The transport potential $\chi$ is 
illustrated in the left column (frames \textit{a}, \textit{f}).
This is derived from the divergence of the convective angular 
momentum transport shown in frames (\textit{b}, \textit{g})
via eq.\ (\ref{eq:poisson}) and helps to elucidate its origin.

In the slowly-rotating regime the angular momentum transport
at mid latitudes is radially inward.  This can be seen from
the nearly horizontal orientation of the $\chi$ contours in 
Fig.\ \ref{fig:chi1}\textit{f}, increasing from top to bottom
(radially inward $\del \chi$).  Since the radial component of
${\bf F}_{RS}$ vanishes at the (impermeable) top and bottom 
boundaries, this implies a divergence of the convective angular
momentum transport in the upper CZ and a convergence in the
lower CZ, as seen in the red and blue layers of 
Fig.\ \ref{fig:chi1}\textit{g}.  

At low latitudes, $\del \chi$ turns equatorward 
(Fig.\ \ref{fig:chi1}\textit{f}), producing 
a convergence of ${\bf F}_{RS}$ at the equator
(Fig.\ \ref{fig:chi1}\textit{g}).  Though this convergence
of the convective angular momentum flux at the equator
tends to establish a solar-like differential rotation,
it does not succeed in doing so.  Rather, it is offset 
by the radially outward transport of angular momentum
by the induced meridional flow (Fig.\ \ref{fig:chi1}\textit{j}).
This induced meridional flow, together with the radially inward 
angular momentum transport at higher latitudes is responsible
for the anti-solar differential rotation profile that is
ultimately achieved.  We discuss this issue further in 
\S\ref{sec:poles}.

This scenario emphasizes the subtle nonlinear, non-local nature 
of the dynamical balances discussed in \S\ref{sec:gp}.  The 
convective angular momentum transport ${\bf F}_{RS}$ is essential 
for establishing the differential rotation $\del \Omega$ but 
it does not uniquely determine the resulting $\Omega$ 
profile.  Rather, ${\bf F}_{RS}$ induces a meridional flow
(mediated by the Coriolis force) that plays an essential
role in establishing the differential rotation.

The role of the convective angular momentum transport in
establishing the meridional circulation is illustrated
in Fig.\ \ref{fig:chi1}\textit{h}.  As expressed by
eq.\ (\ref{eq:gp}), a convergence of ${\bf F}_{RS}$ 
(blue) induces a flow away from the rotation axis
(white arrows) and a divergence of ${\bf F}_{RS}$ 
(red) induces a flow toward the rotation axis (black
arrows).  Mass conservation then requires these
flows to form closed circulation cells, as reflected
by the actual meridional flow pattern shown in 
Fig.\ \ref{fig:chi1}\textit{i}.  Due to the nature
of ${\bf F}_{RS}$ (which can ultimately be traced
to the $\chi$ profile in Fig.\ \ref{fig:chi1}\textit{f}),
the induced meridional flow has a single-celled
nature, dominated by one large cell in each hemisphere
that extends from the equator to a latitude of at
least 70$^\circ$ and spans the entire CZ.  These
cells exhibit poleward flow in the upper CZ and 
equatorward flow in the lower CZ.  The advection
of angular momentum by this induced flow nearly
balances the convective angular momentum transport 
as expressed by eq.\ (\ref{eq:gp}).  This can be seen
by comparing Figs.\ \ref{fig:chi1}\textit{g} and \ref{fig:chi1}\textit{j}.
Small imbalances are due to viscous torques [eq.\ 
(\ref{eq:torque})] and unsteady fluctuations.

Similar arguments hold for the rapidly-rotating regime in the upper
row of Fig.\ (\ref{fig:chi1}).  Here the $\chi$ contours at high
latitudes are again nearly horizontal, signifying a radially inward
angular momentum transport.  However, outside the tangent
cylinder, the angular momentum transport is cylindrically outward, away
from the rotation axis.  This is reflected by the cylindrical nature
of the $\chi$ contours at low latitudes.  There is also an equatorward
contribution to $\del \chi$, particularly at mid-latitudes near the
outer boundary.  This $\chi$ profile produces a Reynolds stress
divergence pattern that reverses sense at high and low latitudes, as
seen in the red and blue regions of Fig.\ref{fig:chi1}\textit{b}.  At
high latitudes the pattern is similar to the high-$Ro$ case, with a
divergence and convergence in the upper and lower CZ respectively.  At
low latitudes this reverses, exhibiting a convergence in the upper CZ
and a divergence in the lower.

The implications of this $\chi$ profile for the meridional circulation
are profound.  Inside the tangent cylinder, a single cell is
established much like the high-$Ro$ case (Figs.\
\ref{fig:chi1}\textit{h},\textit{i}).  However, outside the tangent
cylinder, a multi-cell profile is induced with 2-3 distinct cells
spanning the CZ (Fig.\ \ref{fig:chi1}\textit{d}).  This can be
attributed largely to the cylindrically-outward angular momentum
transport near the equator which induces upward and downward flow in
the upper and lower CZ respectively (Figs.\ \ref{fig:chi1}\textit{c}).
As in the high-$Ro$ regime, the advective angular momentum transport
by this induced meridional flow largely balances the convective
angular momentum transport (Figs.\
\ref{fig:chi1}\textit{b},\textit{e}), with small departures primarily
due to the contribution of viscous diffusion to the zonal force
balance.

The dramatic difference between the convective angular momentum
transport at high and low latitudes can be appreciated by 
considering the structure of the convective flow and how it 
is influenced by rotation.  At high latitudes, convection is
dominated by a quasi-isotropic, interconnected network
of downflow lanes near the surface that break up into more 
isolated lanes and plumes deeper down (Fig.\ 
\ref{solaranti}).  As these downflows
travel downward, the Coriolis force deflects them in a 
prograde direction, inducing a negative correlation between
$v_r^\prime$ and $v_\phi^\prime$ that produces an inward
angular momentum transport (${\bf F}_{RS}$).

At low latitudes, outside the tangent cylinder, the preferred convection
modes for low $Ro$ are {\em banana cells}, columnar convective rolls 
that are aligned with the rotation axis near the equator 
\cite[e.g.,][]{miesc05}.  In simulations with large density stratification
and moderate values of $Ro$, these banana cells are manifested
as a prominent north-south alignment of downflow lanes near the
equator that often trail off eastward at higher latitudes where
they are distorted by the differential rotation
(Fig.\ \ref{solaranti}\textit{b}; see also Miesch et al.\ 2008).

In their simplest manifestation at low $Ro$ and moderate density
stratification, banana cells are approximately aligned with
the rotation axis with little variation in the axial ($z$) 
direction.  The combined action of the spherical geometry,
the density stratification, and positive nonlinear feedback
from the differential rotation they establish, tends to 
produce a systematic tilt such that cylindrically outward
flows ($v_\lambda > 0$, where $\lambda$ is the cylindrical radius) are deflected westward and 
cylindrically inward flows ($v_\lambda < 0$) eastward
\citep{busse02,aurno07}.  This establishes a negative 
$\left<v_\lambda^\prime v_\phi^\prime\right>$ correlation 
that is responsible for the cylindrical orientation of the
$\chi$ contours at low latitudes in Fig.\ \ref{fig:chi1}\textit{a}
(i.e.\ the cylindrically outward angular momentum transport).
The effects of density stratification, the outer spherical
boundary, and moderate (but still low) $Ro$ tend to cause banana 
cells to bend away from the $z$ axis toward a more horizontal
orientation, with axes more parallel the $\theta$ dimension.
This tends to establish positive 
$\left<v_\theta^\prime v_\phi^\prime\right>$ correlations (in the northern
hemisphere, negative in the southern) that
transport angular momentum equatorward \citep{gilma83,glatz84,miesc05}.

Thus, banana cells can account for both the cylindrically outward
angular momentum transport at low latitudes in the low-$Ro$ regime
(Fig.\ \ref{fig:chi1}\textit{a}) and the equatorward angular momentum
transport at low latitudes in the high-$Ro$ regime (Fig.\
\ref{fig:chi1}\textit{f}).  Though the region outside the tangent
cylinder contains the banana cells, a close look at Fig.\
\ref{fig:chi1}\textit{a} indicates that this region does not strictly
delineate the transition between the high and low-latitude regimes for
the angular momentum transport.  Rather, the region of inward
transport at high latitudes shifts equatorward with increasing $Ro$,
as indicated schematically with the solid black lines at mid latitudes
in Figs. \ref{fig:chi1}\textit{c} and \ref{fig:chi1}\textit{h}.

This equatorward shift of the transition between high and low-latitude
behavior can be understood by considering a downflow plume spawned
from the upper boundary layer.  Though such a plume is subject to
a range of forces including pressure gradients, buoyancy, nonlinear 
advection and viscous diffusion, it is instructive to consider 
a ballistic trajectory subject only to the Coriolis force.  Then
a downflow plume with initial velocity $- v_p \uvr$ will be 
deflected in a prograde direction with an initial radius of 
curvature given by
\begin{equation}
r_c = \left\vert\frac{v_p^2}{\pd v_\phi/\pd t}\right\vert = \frac{v_p}{2 \Omega_0 \sin\theta} ~~~.
\end{equation}

The value of $r_c$ relative to $D$, the depth of the CZ, is a measure of
the local, latitude-dependent Rossby number experienced by a downflow
plume:
\begin{equation}\label{eq:loRo}
R_\theta = \frac{r_c}{D} = \frac{v_p}{2 \Omega_0 D \sin\theta}  ~~~.
\end{equation}
Intuitively, eq.\ (\ref{eq:loRo}) can be interpreted based on whether
or not a downflow plume on a ballistic trajectory will traverse the 
CZ before being deflected by the Coriolis force.  If $R_\theta >> 1$ 
then the plume could in principle (in the absence of other forces)
reach the base of the CZ with only a small prograde (positive $\phi$)
deflection.  This Coriolis-induced deflection reflects the tendency
for the plume to conserve its angular momentum conservation and 
it is responsible for the negative $\left<v_r^\prime v_\phi^\prime\right>$ 
correlation that transports angular momentum inward at high latitudes
as seen in Figs.\ \ref{fig:chi1}\textit{a} and \textit{f}.

For $R_\theta << 1$ the plume will not make it to the base of the CZ
(again, considering the simplified case of a ballistic trajectory).
Rather, the convective flow will have to restructure itself in order
to provide the requisite heat transport.  This is the realm of 
banana cells.  This suggests that the transition 
between the two regimes should occurs where $R_\theta \approx 1$.
However, we found above that the actual transition occurs at a 
lower threshold Rossby number of $R_t \sim 0.16$ when $Ro$ is
defined based on the rms velocity and the depth of the CZ
(cf.\ Fig.\ \ref{rossb_domega}). In the present context this 
implies that the transition to the rapid-rotation regime occurs 
when a ballistic plume is deflected well before it reaches the
base of the CZ; $r_c < R_t D$.  This suggests a transition 
colatitude of
\begin{equation}\label{eq:theta0}
\theta_0 \approx \sin^{-1}\left(\frac{v_p}{2 \Omega_0 D}\right)  
\approx \sin^{-1}\left(\frac{Ro}{R_t}\right) ~~~,
\end{equation}
where $Ro$ is the global Rossby number defined in \S\ref{sec:intro}
and where we have assumed that $v_p \sim v_{rms}$.  For the solar-like
case A3 ($Ro = 0.11$), this corresponds to a transition latitude of
50$^\circ$.  This estimate agrees well with the actual transition
latitude indicated in Fig.\ \ref{fig:chi1}$c$.

Equation (\ref{eq:theta0}) should be regarded only as a loose rule 
of thumb for accounting for why the transition between polar and
equatorial regimes for the convective angular momentum transport
shifts equatorward with increasing $Ro$.  It formally breaks
down for $Ro \geq R_t$ where $R_\theta \geq R_t$ at all latitudes.
Yet, even in this regime, banana cells contribute equatorward
angular momentum transport near the equator as seen in 
Fig.\ \ref{fig:chi1}\textit{f}.

In summary, we find that the angular momentum transport by the
convective Reynolds stress, ${\bf F}_{RS}$, plays a central role in
establishing the meridional flow profiles in our simulations.  In
particular, we find a transition from single-celled to multi-celled
meridional circulation profiles in the high and low $Ro$ regimes that
is directly linked to a change in the nature of ${\bf F}_{RS}$.

This conclusion is supported by the recent 
high-resolution simulations of \cite{hotta14b}.  There they find 
that inward angular momentum transport by small-scale convection
in the surface layers induces a poleward meridional flow as
envisioned by MH11.  As in our simulations, this is a clear 
demonstration of MC profiles established by gyroscopic pumping
(see also \S\ref{sec:nssl}).

%%%%%%%%%%%%%%%%%%%%%%%%%%%%%%%%%%%%%%%%%%%%%%%%%%%%%%%%%%%%

\subsection{The Solar-Anti-solar Transition}\label{sec:poles}

In \S\ref{sec:amom} we demonstrated that the convective angular 
momentum transport ${\bf F}_{RS}$, plays a central role in regulating
the MC profile in both the high and the low $Ro$ regimes.  
In this section we demonstrate that ${\bf F}_{RS}$ also regulates
the transition between the two regimes and that the induced MC
plays a key role in this transition.

We focus primarily on the anti-solar (high $Ro$) regime where
the convective angular momentum transport is radially inward
at most latitudes (Fig.\ \ref{fig:chi1}\textit{f}).  As discussed 
in \S\ref{sec:amom} and as shown in Fig.\ \ref{fig:chi1}\textit{i}, 
this induces a single-celled MC that is counter-clockwise in 
the northern hemisphere (NH) and clockwise in the southern 
hemisphere (SH).  This MC transports angular momentum that offsets 
the convective angular momentum transport (Fig.\ \ref{fig:chi1}\textit{j}).

However, the MC induced by gyroscopic pumping also transports 
entropy.  Since the stratification of the CZ is superadiabatic
($\pd S/\pd r < 0$), the upflows and downflows at low and high
latitudes tend to establish an equatorward entropy gradient
($\pd S/\pd \theta$ positive in the NH and negative in the
SH).  This gives rise to a baroclinic forcing that further
enhances the MC.  Baroclinicity thus provides
a positive feedback that amplifies the MC established by
gyroscopic pumping.  Essentially, the mechanical forcing
due to ${\bf F}_{RS}$ triggers an axisymmetric convection
instability that feeds back on the mean flows by redistributing
both angular momentum and entropy.  This axisymmetric mode is superposed on the preferred 
non-axisymmetric modes for convection in rotating spherical 
shells \citep[e.g.]{miesc09}.  Though rotation plays
an essential role in exciting this axisymmetric convective mode 
through the Coriolis-induced Reynolds stress, it may have some 
bearing on similar large-scale circulations observed in laboratory 
experiments and simulations of non-rotating turbulent Rayleigh-B\'enard 
convection \citep{ahler09}.

\begin{figure}
\centerline{\epsfig{file=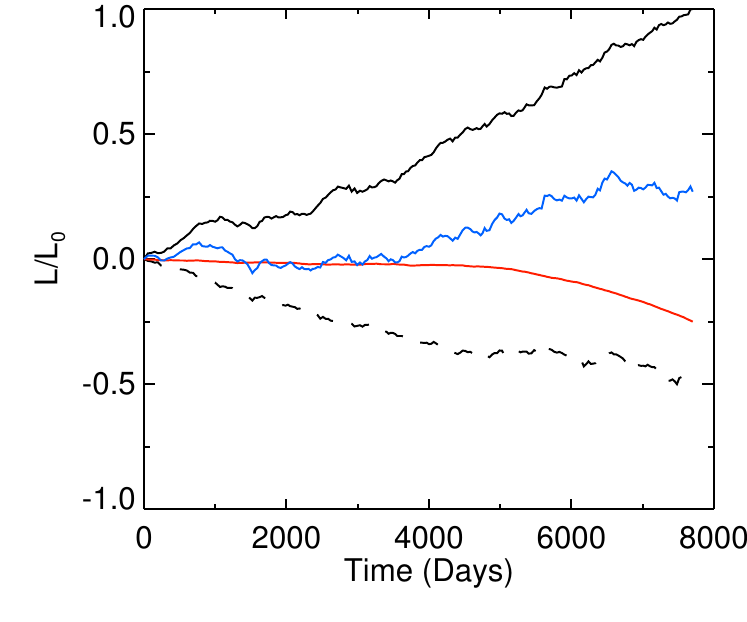,width=\linewidth}}
\caption{\label{amom_time}Polar-spinup in anti-solar case C1.  Total angular momentum transported poleward across 60$^\circ$ N latitude, as a function of time, is shown in blue.  The individual contributions to this transport by the meridional circulations (black line), convective Reynolds Stresses (dashed line), and viscous stresses (red line) are indicated as well.  Polar spin-up in case C1 arises from an imbalance in transport by Reynolds stresses, which work to spin up the equator, and transport by meridional circulations, which work to spin up the poles.  Equilibration occurs around day 6000 as viscous stresses associated with strong rotational shear become strong enough to help counter the effect of meridional circulation.  } 
\end{figure}

This is demonstrated in Fig.\ \ref{amom_time}, which illustrates
how the anti-solar differential rotation in Case C1 is established.
Shown in blue is the time evolution of $L$, which is the angular momentum 
of the north polar region relative to the rotating reference frame: 
\begin{equation}\label{eq:Lpole}
L_p = \int_{r_1}^{r_2} \int_0^{\theta_p} \int_0^{2\pi} \overline{\rho} r^3 \sin^2\theta v_\phi dr d\theta d\phi ~~~,
\end{equation}
where $\theta_0 = \pi/6$, corresponding to a latitude of $60^\circ$.  In 
Fig.\ \ref{amom_time}, $L$ is normalized by $L_0$, the total amount transport poleward by meridional circulations over this interval.
The stress-free boundaries exert no torques so the total angular momentum
[obtained by replacing $\theta_0$ with $\pi$ in eq.\ (\ref{eq:Lpole})] is 
conserved.  Thus, the time evolution results from the integrated angular
momentum flux across a latitude of 60$^\circ$.  This flux includes
contributions from the convective Reynolds stress, the meridional circulation,
and the viscous diffusion, represented in Fig.\ \ref{amom_time} by a dashed
line, a solid black line, and a red line respectively.

The first thing to note in Fig.\ \ref{amom_time} is that the
convective angular momentum transport is not strictly radial.  Rather,
it includes a weak equatorward component that acts to spin up the
equator even at mid latitudes.  Over the first 4000 days of evolution,
this is approximately balanced by the meridional circulation, which
transports angular momentum poleward.  However, eventually the MC
overwhelms the Reynolds stress, spinning up the poles until viscous
diffusion sets in to help limit the growth of the cyclonic polar
vortex.

Thus, it is the meridional circulation that ultimately establishes the
anti-solar differential rotation, not the Reynolds stress directly, 
although the latter induces the former.  The transition
from solar to anti-solar differential rotation is triggered by the
inward angular momentum transport but mediated by the strong, 
single-celled MC.  This stems from the tendency for the MC
to homogenize angular momentum, spinning up the poles relative to
the equator (Paper 1) and is consistent with the poleward 
angular momentum transport by the MC reported in many global
convection simulations \citep{brun02,miesc08,kapyl11}.

The role of the MC in establishing anti-solar DR 
profiles has previously been emphasized in mean-field models by
\cite{kitch04} \citep[see also][]{rudig07}.  However, in these
models the MC is not established by convective transport. 
Rather, other processes such as the suppression of high-latitude
convection by polar starspots or tidal forcing from a binary
companion induce baroclinic torques that in turn induce global 
circulations.  By contrast, in our convection models, the MC is 
established by a radially inward convective angular momentum 
transport and reinforced self-consistently by baroclinicity.

%%%%%%%%%%%%%%%%%%%%%%%%%%%%%%%%%%%%%%%%%%%%%%%%5
%%%     Section 7
%MMMMMMMMMMMMMMMMMMMMMMMMMMMMMMMMMMMMMMMMMMMMMMMMMMMMMMMMMMMMMMMMMMMMMM

\section{Convective Heat Transport in Solar and Anti-Solar Cases}
\label{sec:heat}

In \S\ref{sec:gp}-\S\ref{sec:amom} we argued that it is the 
convective angular momentum transport that ultimately determines the gross
differential rotation contrast $\Delta \Omega$ as well as the MC profile.
However, we have also seen that baroclinicity can play an important role
as well, both in establishing anti-solar DR profiles by spinning up the 
poles (\S\ref{sec:poles}) and in shaping the $\Omega$ contours through
thermal wind balance, eq.\ (\ref{eq:twb}).  Though most convection and
mean-field models agree that TWB is responsible for the conical $\Omega$
profiles in the solar CZ (\S\ref{sec:gp}), the origin of this baroclinicity
is not well understood.   Early models attributed it to the influence
of rotation on convective heat transport \citep{kitch95,robin01,brun02}
but thermal coupling to the tachocline may also play an important
role \citep{rempe05,miesc06}.  Here we consider the former by investigating
the inhomogeneous nature of the convective heat flux in the solar and
anti-solar regimes.

\begin{figure}
\centerline{\epsfig{file=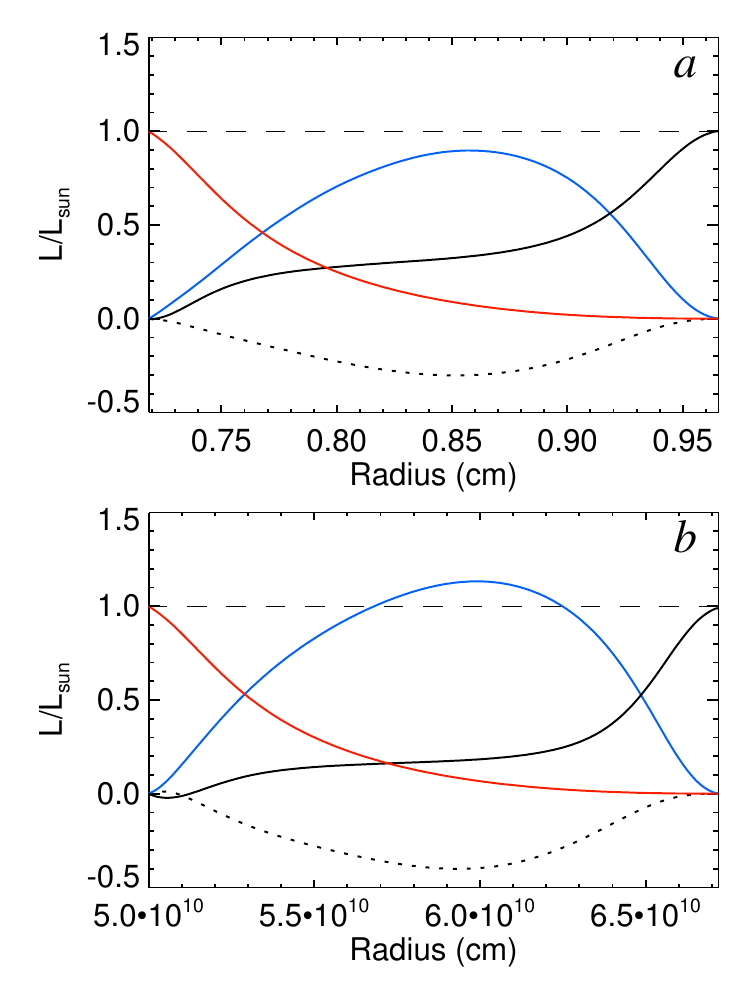,width=\linewidth}}
\caption{\label{flux_bal} Energy flux balance in cases B1 (\textit{a}; our most turbulent solar-like case) and C1 (\textit{b}; the most strongly driven anti-solar case).  Fluxes have been integrated over the shell to yield a luminosity.  Enthalpy flux is shown in blue, radiative flux in red, conductive flux in solid black, and inward kinetic energy flux in dotted black.  A dashed line has been plotted at unity for reference.  Convection tends to transport additional heat in the anti-solar case relative to the solar case, compensated for by an increased inward kinetic energy flux.}
\end{figure}

\begin{figure}
\centerline{\epsfig{file=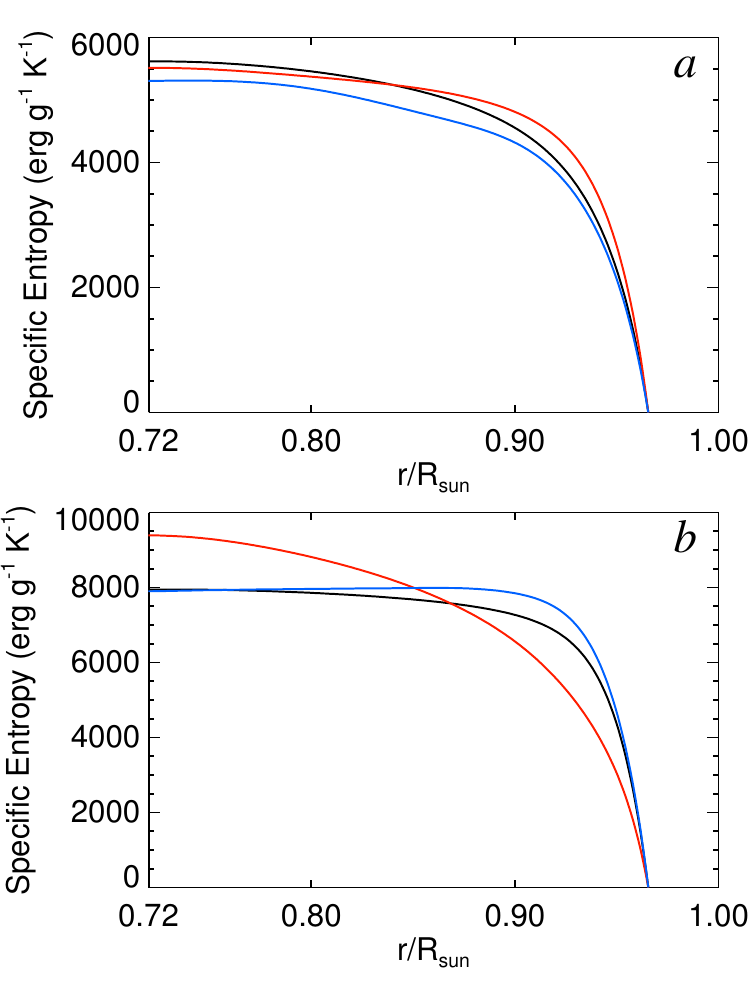,width=\linewidth}}
\caption{\label{entropy_cuts}Variation of entropy in the (\textit{a}) low Rossby number (case A3) and (\textit{b}) high Rossby number (case C1) regimes.  The spherically symmetric component of the entropy perturbations is plotted in black for each case.  Cuts at the equator and a latitude of 85$^{\circ}$ have been plotted in blue and red respectively.  Variations about the spherically symmetric mean have been amplified by a factor of 3 for visibility.  Convection is more efficient at high latitudes in the low Rossby number regime, signified by the steeper entropy gradient near the surface and the nearly adiabatic interior in those regions.  For the low Rossby number regime of (\textit{b}), convection is more efficient at low latitudes.}
\end{figure}

\begin{figure}
\centerline{\epsfig{file=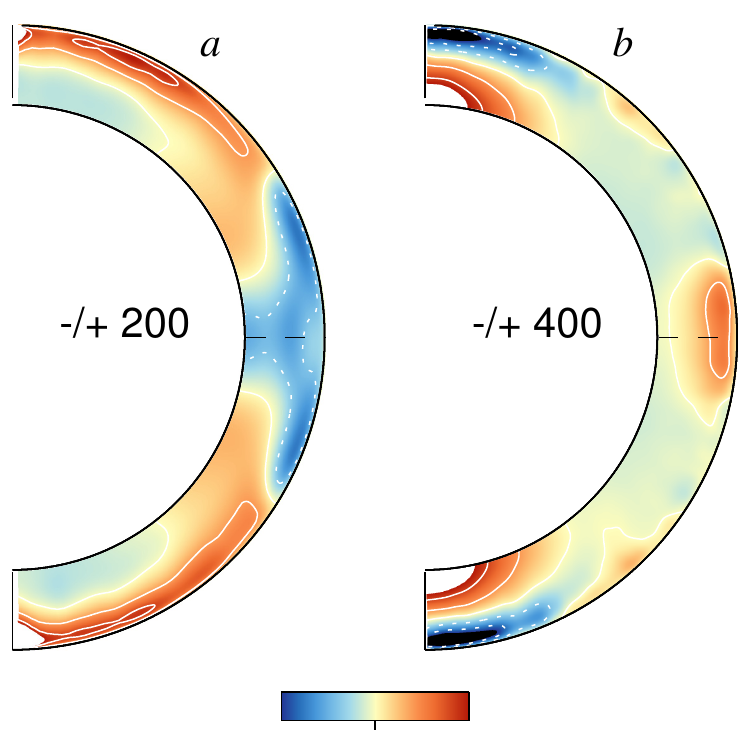,width=\linewidth}}
\caption{\label{entropy_azavg}Longitudinally averaged entropy perturbations, with spherically symmetric mean subtracted at each radius, for solar-like case A3 and anti-solar case C1.  The solar-like cases tend to develop polar regions that warm with respect to the equator.  By contrast, the anti-solar case exhibits a warm equator.  These longitudinal averages have also been time-averaged over a period of roughly 200 days.} 
\end{figure}

A sense of the energy flux balance in the rapidly-rotating and slowly-rotating regime may be found by examining Figure \ref{flux_bal}.  Plotted in Figure \ref{flux_bal} are the convective enthalpy flux $F_e$, the radiative heat flux $F_r$, the kinetic energy flux $F_k$, and the conductive heat flux $F_c$, defined as
\begin{equation}
\label{eq:enthalpy}
F_e = \avg{\rho}c_p\overline{v_rT'},
\end{equation}
\begin{equation}
\label{eq:frad}
F_r=-\kappa_r\avg{\rho}c_p\frac{d\overline{T}}{dr},
\end{equation}
\begin{equation}
\label{eq:fke}
F_k = \frac{1}{2}\avg{\rho}\overline{v_r\vec{v}^2},
\end{equation}
and
\begin{equation}
\label{eq:fcond}
F_c = -\avg{\rho}\avg{T}\kappa\frac{\partial \avg{S}}{\partial r}.
\end{equation}
Each of these fluxes has been integrated over horizontal surfaces and normalized by the solar luminosity.  The outward convective transport of heat and inward transport of kinetic energy dominate the flux balance at mid-convection zone in either system.  The rapidly rotating system B1 (Figure \ref{flux_bal}\textit{a}) has somewhat less efficient convection than the slowly-rotating system (Figure \ref{flux_bal}\textit{b}), and thus possesses a stronger contribution from conduction at mid-convection zone.  This behavior is typical of other rapidly rotating systems; namely a strong rotational constraint tends to inhibit the convective efficiency.  The latitudinal averaging that has gone into Figure \ref{flux_bal} belies significant differences in the convective state achieved by these systems, however.   

The convective transport of heat achieved in either the solar or anti-solar regimes is highly anisotropic latitudinally, with the nature of that anisotropy depending on which rotational state the convection is in.  Due to the impenetrable outer boundary present in these systems, transport of heat across the upper boundary is accomplished by conduction.  It is in this region where the super-adiabaticity of the system is established, resulting from the development of a conductive boundary layer.  As we have chosen to fix the value of entropy at the upper boundary, the entropy gradient there is free to vary in time, and also in latitude.  Fig. \ref{entropy_cuts} illustrates the typical thermal background realized in the rapidly rotating (Fig. \ref{entropy_cuts}\textit{a}; case A3) and slowly-rotating (Fig. \ref{entropy_cuts}\textit{b}; case C1) regimes. The spherically symmetric entropy profile is plotted in black alongside cuts of the axisymmetric entropy profiles at the equator (red) and 85$^{\circ}$ (blue).  At high latitudes, super-adiabaticity in the entropy gradient is confined to a thin boundary layer for the solar-like case, implying efficient convection.  The situation is the opposite at low latitudes, where much of the convection zone is super-adiabatic, albeit with a lower overall adiabat.  This situation, with efficient polar convection and inefficient equatorial convection is typical of the rapidly-rotating regime and arises from the prominent role played by the Coriolis force, which tends to inhibit radial downdrafts more efficiently at the equator than at the poles.  

The slowly-rotating regime exhibits an even more pronounced dichotomy between pole and equator, but inverted.  Here convection is most efficient at the equator, with a super-adiabatic stratification at high latitudes (Fig.\ \ref{entropy_cuts}\textit{b}).  Here the convection is suppressed by the strong prograde polar vortex that is characteristic of the anti-solar regime.  Down-welling plumes are quickly deflected in the prograde direction by the strong rotational shear and the cylindrically outward Coriolis force.  The polar vortex also acts as a transport barrier for meridional momentum, suppressing the meridional flow as can be seen by examining the C cases in Fig.\ \ref{circ_grid}.   

The thermal anisotropies realized in the two different regimes have implications for the establishment of meridional circulation as well.  Profiles of the axisymmetric entropy perturbations are plotted in Fig. \ref{entropy_azavg} where, in order to emphasize the latitudinal variation of the entropy perturbations, we have subtracted out the spherically symmetric mean.  In case A3 (Fig. \ref{entropy_azavg}\textit{a}), the polar regions are warm with respect to the equator.  In isolation, this effect alone would drive a circulation that is \textit{equatorward} at the surface.  This is not the nature of the circulation realized in the rapidly rotating regime.  In these simulations, the warm poles thus serves to \textit{suppress} the meridional circulations realized.  The meridional circulations in the rapidly-rotating regime are thus \textit{not driven thermally}.  The same can be said of the Sun (Paper 1).

The situation at the surface of C1 is reversed, however, and the effect is to augment any poleward circulations already established in the near-surface layers by gyroscopic pumping. This reflects the excitation of an axisymmetric convective mode as discussed in \S\ref{sec:poles}.  Here baroclinicity is unable to achieve thermal wind balance and viscous forces are needed to establish equilibrium.  In actual stars where viscous forces are negligible, this role may be played by magnetic fields, which would inhibit the development of a polar vortex.

\section{Application to the Sun and stars}\label{sec:sas}

\subsection{What Regime is the Sun in?}\label{sec:sun}
What do our results imply about the meridional circulation in 
the Sun?  Is it single-celled or multi-celled?  Since its rotation
profile is solar-like, by definition ($\Delta \Omega > 0$), we
might readily place it in the rapidly-rotating regime and we
might therefore expect that it should have a multi-celled profile.  However,
the situation is more subtle than this simple categorization
suggests.  The difficulty lies with the conical nature of the Sun's $\Omega$
profile.

There is a clear difference between the ``solar-like'' differential rotation profiles
in the low-Rossby number simulations presented in \S\ref{sec:regimes} and the actual
internal rotation profile of the Sun inferred from helioseismology.
Like the Sun, the simulations in the rapidly rotating regime exhibit a
fast equator and slow pole ($\Delta \Omega > 0$).  However, unlike the
Sun, the simulations exhibit rotation profiles that are approximately
cylindrical, such that $\del \Omega$ is directed away from the
rotation axis.  By contrast, helioseismic rotational inversions
suggest that rotation profiles are more conical in nature, such that 
$\del \Omega$ is primarily equatorward, with nearly radial $\Omega$
isosurfaces at mid-latitudes as shown in Fig.\ \ref{fig:sunfig}\textit{a}
\citep[see also][]{thomp03,howe09}.  The presence of these conical 
profiles may signify that the Sun is near the transition 
between the two mean flow regimes.

To illustrate why this is the case, we must first understand
why the $\Omega$ profile in the Sun is conical rather than
cylindrical.  As discussed in section \ref{sec:gp}, a diverse
variety of modeling efforts including convection simulations,
mean-field models, and theoretical arguments have converged on a common
explanation. They attribute the conical nature of the solar rotation
profile to thermal gradients, which contribute a baroclinic forcing to
the zonal vorticity equation that can effectively offset meridional
flows induced by the inertia of the differential rotation that would
otherwise establish cylindrical profiles in accordance with the
Taylor-Proudman theorem.  This robust balance between baroclinicity
and inertia is expressed by the thermal wind equation (\ref{eq:twb}),
which implies
\begin{equation}\label{twb2}
\Omega \frac{\pd \Omega}{\pd z} \propto \Delta S
\end{equation}
where $\Delta S$ is the entropy difference between pole and equator.
How the differential rotation profile changes
as the global rotation rate $\Omega$ is increased ultimately depends on the scaling of $\Delta S$ with rotation rate.

The mean-field models of
\cite{hotta11}, which include thermal coupling to the tachocline, suggest that $\Delta S$ does not scale steeply enough with $\Omega$ to maintain a conical differential rotation for fast rotators.  These models exhibit an increasingly cylindrical differential rotation
profile as $\Omega$ is increased.  Similarly, simulations of G-type stars 
by \cite{brown08} and recent mean-field models by \cite{kuker11}
also show nearly cylindrical alignment at rapid rotation rates.
However, in mean-field models \citep{kuker05} and 3-D simulations 
\citep{augus12} of convection in F-type stars with thinner convection 
zones, thermal gradients are able to scale with rotation gradients,
maintaining a significant conical orientation over nearly an
order of magnitude in $\Omega$.

If $\Omega$ profiles do become more cylindrical at faster rotation
rates, then the conical $\Omega$ profile of the Sun suggests that
it is in a low-Rossby number (rapidly-rotating) regime but perhaps
only marginally.  What does this imply about the meridional flow
profile?  

In section \ref{sec:interp} we argued that the multi-cell meridional
flow profile at low latitudes in rapid rotators can be attributed 
to cylindrically outward angular momentum transport by columnar
convective modes (``banana cells'').  This tends to produce not
only a multi-cell circulation profile but also a cylindrical
$\Omega$ profile with a prominent $\pd \Omega / \pd r$ at the 
equator (see Fig.\ \ref{spindown}). 

The solar rotation profile exhibits a weak $\pd \Omega/\pd r$ 
near the equator with the same sense (positive) as in the 
rapidly-rotating solar models.  This may indicate that some
cylindrically outward angular momentum transport is indeed
occurring.  However, the angular velocity contrast $\Delta \Omega$
across the CZ at the equator (10-20 nHz) is small compared 
to the contrast between the equator and mid-latitudes 
(50-60 nHz), suggesting that the convective angular momentum 
transport may be more equatorward rather than cylindrically
outward.

We do indeed find equatorward angular momentum transport
by convection at low latitudes in the slowly rotating
regime, as demonstrated by the nearly radial $\chi$ 
contours adjacent to the equator in Fig. \ref{fig:chi1}\textit{f}.  This
convergence of angular momentum flux at the equator
tends to accelerate the rotation nearly uniformly
throughout the CZ.  Furthermore, through the mechanism
of gyroscopic pumping, 
it establishes a single-cell meridional circulation
profile, with radially outward flow at the equator
that veers poleward at the surface.  

In the slowly rotating regime, the convergence of the
convective angular momentum flux at the equator is not 
sufficient to establish a solar-like differential rotation
profile.  Rather, the radially inward angular momentum
transport at higher latitudes and the meridional flow
it establishes efficiently spin up the poles, producing
an anti-solar profile.  However, there might be situations
in which equatorward angular momentum transport (as opposed
to cylindrically outward) can sustain a solar-like 
$\Omega$ profile.

One such example is our transitional case 
B0.5 discussed in section \ref{sec:transition}.  As shown
in Figure \ref{turb_solar}, this case exhibits a solar-like 
$\Omega$ profile ($\Delta \Omega > 0$) with a single-cell 
meridional flow.  The convective angular momentum transport
in this case is shown in Fig.\ \ref{fig:chi2}.  As in the
anti-solar regime (Fig.\ \ref{fig:chi1}\textit{f}), it is 
radially inward at high latitudes and equatorward at low
latitudes.  Yet here the $\Omega$ profile is solar-like.

Another way to approach this issue is to turn the
problem around, as described by \citet{miesc05}.  Using the 
solar rotation profile inferred from helioseismology, 
we can ask what Reynolds stress would be needed to maintain
it for a given meridional flow profile.  Fig.\ \ref{fig:sunfig} 
shows the result for a single-celled profile.  In other words,
Fig.\ \ref{fig:sunfig}\textit{d} shows the Reynolds stress 
that would be needed to establish a single-celled MC profile
in the Sun, expressed in terms of the transport potential $\chi$.  
This looks notably similar to the $\chi$ contours in the transitional
case B0.5, shown in Fig.\ \ref{fig:chi2}\textit{a}.  The latter
are more radial at low latitudes but this is sensitive to the
precise form of the MC; what matters here is the qualitative
similarity.  Also, the $\Omega$ profile in case B0.5 is not 
completely analogous to the Sun in that it is cylindrical 
rather than conical.  However, this discrepancy may have to 
do with the absence of a tachocline, as discussed in \S\ref{sec:tacho}.

Our results suggest that the Sun may be near the transition
between the fast and slow rotating regimes.  This is consistent with
previous estimates of the Rossby number of the Sun, which lie just below 
unity (Paper 1).  We predict that G-type stars rotating faster than the 
Sun should have cylindrical $\Omega$ profiles with multi-cell meridional
circulation profiles and those rotating much slower than the Sun should have
anti-solar $\Omega$ profiles and single-cell meridional flow profiles.
However, the Sun is close enough to the transition that we cannot 
definitively say from these simulations alone whether it has a 
multi-cell or single-cell profile.  

It is interesting to note that if the mean flows
of the Sun are in a transitional regime and thus atypical relative
to other stars with faster or slower rotation rates, then this should
be reflected in its observed patterns of magnetic activity.  Both the 
differential and the meridional circulation play an essential role in most 
recent dynamo models of the solar cycle \citep[e.g.][]{charb10}.  If these 
are atypical, then the solar cycle should be too.  Although data on stellar
cycles is limited and sometimes ambiguous, there is some evidence that
the period of the solar cycle relative to its rotation rate is indeed 
atypical, lying between the active and inactive dynamo branches described
by \citet{bohm07}.

Further insight must come from continuing solar and stellar
observations, helioseismic inversions, and from the consideration of
factors such as boundary conditions and magnetism, which can
influence the solar/anti-solar transition and the structure of
mean flows.  We address these issues in the following subsections.

\begin{figure*}
\centerline{\epsfig{file=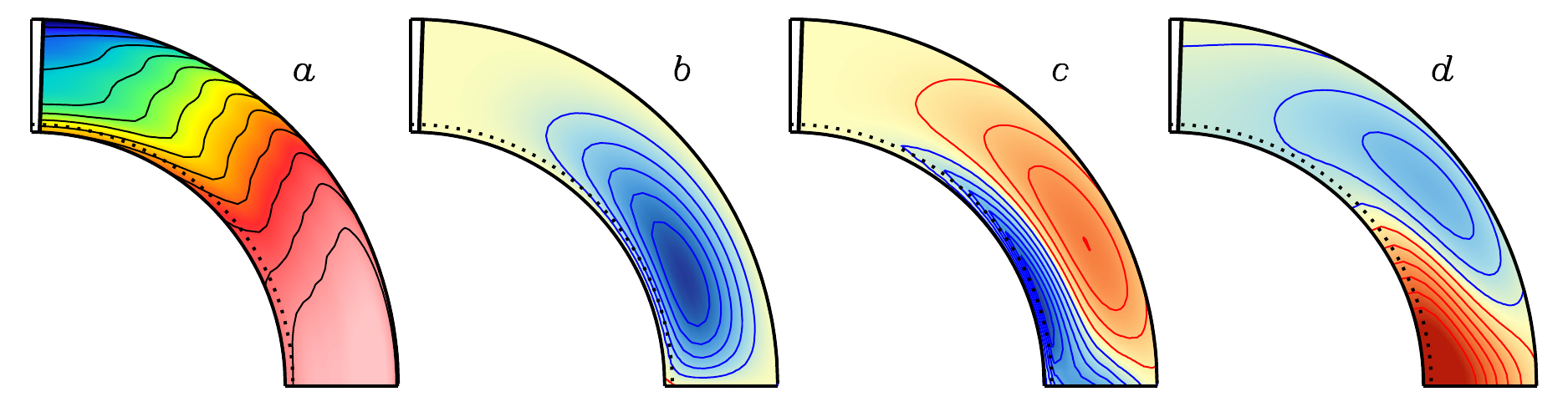,width=\textwidth}}
\caption{\label{fig:sunfig}Inferred angular momentum transport for
the Sun, assuming a single-cell meridional circulation profile
\citep[after][]{miesc05}. ($a$) Angular velocity $\Omega$ based on
RLS inversions of GONG data for four non-overlapping intervals
in 1996, provided by R.\ Howe \citep{howe00,schou02}. ($b$)
Hypothetical single-celled meridional circulation profile,
shown in terms of the streamlines of the mass flux, used in
the flux-transport dynamo models of \citet{dikpa11}. Blue
denotes counter-clockwise circulation; the amplitude is 
arbitrary for our purposes.  ($c$) Mean axial torque 
required to sustain these mean flows, from equation (\ref{eq:gp}): 
${\cal F} = \rh \left<\vv_m\right> \bdot \del {\cal L}$.
Red denotes a divergence of the angular momentum flux 
(${\cal F} < 0$) and blue denotes convergence.
($d$) Corresponding $\chi$ contours, as in Fig. \ref{fig:chi1}.}
\end{figure*}

\begin{figure}
\centerline{\epsfig{file=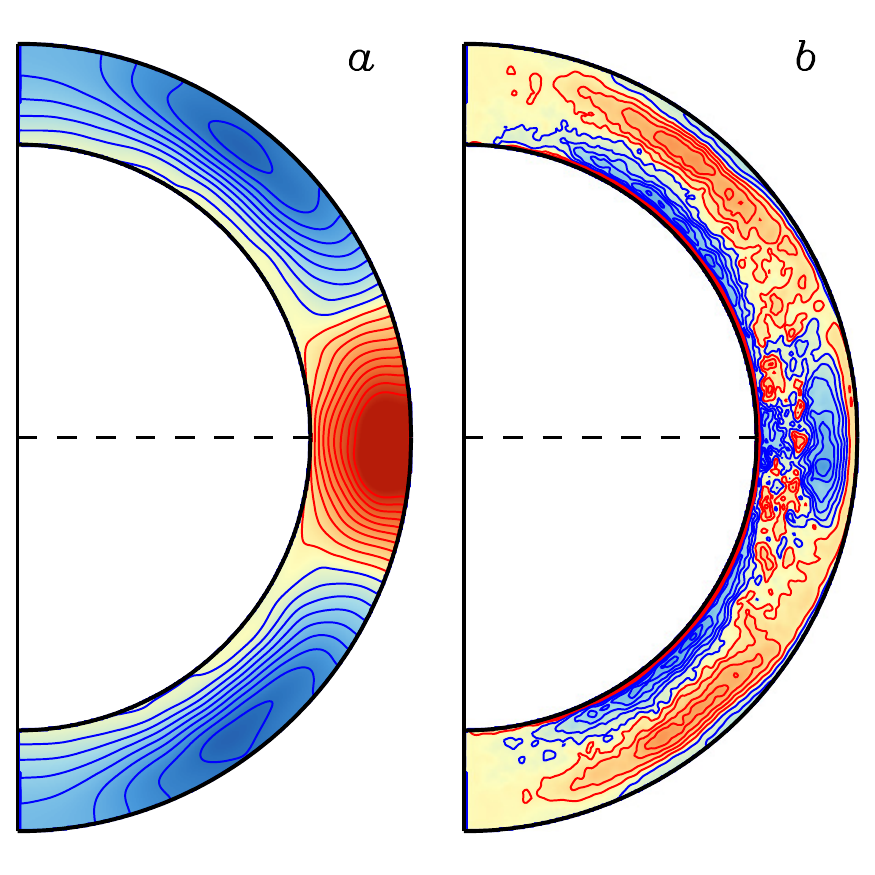,width=\linewidth}}
\caption{\label{fig:chi2} Convective angular momentum transport in the 
transitional case, B0.5.  Shown are (\textit{a}) the transport potential 
$\chi$ and (\textit{b}) the negative divergence of ${\bf F}_{RS}$ (cf.\ 
Fig. \ref{fig:chi1}, frames \textit{a}, \textit{b}, \textit{f}, and 
\textit{g}).  Color tables are as in Fig.\ \ref{fig:chi1} with saturation 
levels of (\textit{a}) 1.4$\times 10^{26}$ g cm s$^{-2}$ and 
(\textit{b}) 2.1 $\times 10^7$ g cm$^{-1}$ s$^{-2}$.}
\end{figure}

\subsection{The Near-Surface Shear Layer}\label{sec:nssl}
The simulations reported on here lack a notable component of the solar convection zone: a near-surface shear layer (NSSL).   This region of shear, located within the outer
5\% of the solar interior, is characterized by a decrease in angular velocity with radius ($\pd \Omega / \pd r < 0$) of about 2-4\%
\citep{thomp03,howe09}.  Our simulations lack an NSSL comparable to the Sun likely because of the simplified upper boundary condition and the lack of
small-scale surface convection (recall from \S\ref{sec:experiments} that 
our simulations only extend to 0.965$R$).  However, the insight gained
from these simulations as discussed in \S\ref{sec:interp} can help guide
the interpretation of helioseismic inversions.

As discussed by MH11, the existence of the
solar NSSL is often attributed to the inward transport of angular momentum
by small-scale convection in the solar surface layers where the effective
Rossby number is high.  The phenomenon of gyroscopic pumping links the
differential rotation and meridional flow in the NSSL and can, at
least in part, account for the poleward flow near the solar surface.
In particular, the inward angular momentum transport by solar
surface convection gives a retrograde torque (flux divergence) 
that sustains the poleward flow through equation (\ref{eq:gp}).
Much of this inward angular momentum transport likely arises 
from convective plumes that are driven by photospheric cooling 
and that are deflected in a retrograde direction by the Coriolis
force as they propagate downward, as described in \S\ref{sec:amom}.
If these plumes penetrate to a radius of $r_p$, then one would
expect a requisite convergence of the angular momentum flux
which would then induce an equatorward flow near $r_p$ as 
expressed by the gyroscopic pumping equation (\ref{eq:gp}).

Thus, the recent detection of a shallow 
equatorward return flow at $r \sim $0.91-0.93$R$ may reflect
the penetration depth of convective plumes that originate in 
the photosphere.  This is consistent with the interpretation
of \citet{hatha12b} who attribute the shallow return flow to 
the presence of the NSSL and the Reynolds stress associated with
supergranular convection.  Furthermore, this may account for why 
the shallow return flows reported by \citet{hatha12b} and 
\citet{zhao13} reverse direction along surfaces that are nearly 
horizontal while the reversal surfaces in our multi-celled
low-Rossby number simulations are more cylindrical (see
Fig.\ \ref{spindown}$f$).  It will be interesting to see if 
future inversions show any signs of cylindrical alignment
in deeper layers where the Rossby number is low.

More generally, helioseismic inversions that focus on the NSSL region
may reflect local dynamics that have little bearing on the
deeper dynamics responsible for maintaining the solar-like
differential rotation ($\Delta \Omega > 0$) and, by extension,
the deep meridional flow.  Since the process of gyroscopic 
pumping is intrinsically nonlocal (MH11), the presence of the 
NSSL may influence the meridional circulation profile throughout 
the convection zone.  However, the lower CZ is certainly not 
passive; this much is clear from the nature of the $\Omega$
profile, which is almost certainly not maintained solely by 
high-Rossby number surface convection.  Thus, even if there 
is a shallow return flow above $0.9 R$, this tells us little 
about the mean circulation at the base of the convection zone, 
which is most relevant for solar and stellar dynamo models 
\citep[e.g.][]{charb10}.

\subsection{The Role of the Tachocline}\label{sec:tacho}

Much as the upper boundary layer of the CZ
might influence the meridional circulation profile of the
solar convection zone, so too might the lower boundary layer, where the transition to a sub-adiabatic radiative zone occurs.  In \S\ref{sec:poles} we attributed the spin-up of the poles 
in the high Rossby number regime to the poleward 
advection of angular momentum by the extended single-cell 
meridional circulation.  The circulation was attributed to
gyroscopic pumping induced by an inward angular momentum 
transport but positive feedback from an equatorward entropy
gradient amplified it, quickly establishing an anti-solar
$\Omega$ profile.  Balance was only achieved when the
shear became large enough for subgrid-scale diffusion 
to suppress further growth.

In actual stars where (microscopic) diffusion is negligible, 
the balance may be achieved quite differently.  For instance, thermal coupling to the tachocline can
induce a baroclinicity throughout the CZ that helps 
warm the poles and thus suppresses the formation of a counter-clockwise
(clockwise) circulation cell in the northern (southern) hemisphere
\citep{rempe05,miesc06,brun11}.  This baroclinicity is primarily due to 
the downward circulation at high-latitudes which burrows down into the 
steep sub-adiabatic stratification of the overshoot region and 
radiative interior.  This downward advection of relatively low entropy
fluid from the CZ induces a strong poleward entropy gradient in 
the tachocline that establishes an approximate thermal wind balance 
with the differential rotation as expressed in eq.\ (\ref{eq:twb}).
Convective heat transport may then transmit this strong poleward
entropy gradient vertically, back up into the CZ where it could
mitigate the equatorward entropy gradient that helps drive the
runaway convective mode discussed in \S\ref{sec:poles}.

In addition to maintaining conical $\Omega$ profiles throughout the CZ, 
this mechanism of thermal coupling may also
suppress the strong CCW meridional circulation cell
that helps to regulate the transition between solar
and anti-solar rotation regimes.  In short, the 
same mechanism that maintains the conical $\Omega$
profile in the Sun might also help maintain the
fast equator and slow pole that defines the solar-like 
profile ($\Delta \Omega > 0$), shifting the 
anti-solar transition ($\Delta \Omega > 0$) toward
lower Rossby number.  The 
conical $\Omega$ profile of the Sun may thus signify 
that is is close to the solar/anti-solar transition, 
as discussed in \S\ref{sec:sun}.  We are currently
investigating these issues with convection simulations 
that include a subadiabatic overshoot region and 
radiative zone, to be described in future papers.

\subsection{Magnetism and Mean-Flow Regimes}\label{sec:mag}

Another issue that warrants further research
is the role of magnetism.  Magnetism may suppress shear and thus assume the role of SGS
diffusion in the limit of low microscopic diffusivity \citep[e.g.][]{nelso13}.  By suppressing shear, the Lorentz force (and in particular the Maxwell
stress component) tends to partially
offset the zonal components of the Reynolds
stress \citep{brun04,nelso13}.  This has the effect
of reducing the net axial torque ${\cal F}$ in 
equation (\ref{eq:gp}), which in turn can reduce
the amplitude of the meridional flow.  This might
account for why the presence of the magnetic
field was found to shift the solar/anti-solar 
transition toward lower Rossby number in 
recent global convection simulations by \citet{fanfang14}.  By suppressing
the meridional flow, magnetism may inhibit the
polar spin-up associated with the anti-solar 
regime.

The inhomogeneous spatial distribution of magnetic
stresses can also suppress or alter meridional flow patterns.  To illustrate the potential consequences of this, consider 
the rapidly-rotating regime, where the cylindrically outward 
angular momentum transport at low latitudes maintains a
multi-cell meridional flow as shown in Fig. \ref{fig:chi1}\textit{c}.  The 
region of divergence near the equator (red) is cylindrically
aligned but its amplitude decreases with increasing distance
from the equatorial plane.  It is this axial variation in the Reynolds stress (parallel to the rotation axis) that helps establish the multi-cellular circulation profile.  In recent simulations of convection in F-type stars, 
\citet{augus13} find that the Lorentz force exerts a
retrograde torque (flux divergence) at mid-latitudes
that supplements the lower-latitude Reynolds stress, and diminishes its axial variation.  The resulting
meridional circulation at the equator in the lower CZ
shifts from inward to outward, though the overall profile
remains multi-celled.  The presence of magnetism is thus likely to effect the detailed meridional flow profiles and the precise
value of the transitional Rossby number between mean flow regimes, 
but it should not change our principle conclusion that rapid rotators
should exhibit multi-cell meridional flow profiles while slow
rotators should exhibit single-cell profiles.

\section{Summary}\label{sec:summary}
%MMMMMMMMMMMMMMMMMMMMMMMMMMMMMMMMMMMMMMMMMMMMMMMMMM
In this paper we report on a series of solar convection simulations spanning a range of 
rotational influences, as quantified by the Rossby number $Ro$.  As in previous work 
\citep{gasti13,gasti14, guerr13, kapyl14}, we find two distinct dynamical regimes of differential rotation (DR). 
One regime is characterized by solar-like DR ($\Delta \Omega > 0$) at low Rossby number ($Ro$) and the other by anti-solar 
DR ($\Delta \Omega < 0$) at high $Ro$, where $\Delta \Omega = \Omega_{eq}-\Omega_{pole}$ is the pole-to-equator rotational contrast.  Our transitional
value of $Ro \sim 0.16$ is lower than that reported in previous studies
but we attribute this to a difference in how $Ro$ is defined.
The most novel aspect of the present work is a focus on the meridional circulation (MC) in the two dynamical regimes, including how it is established and maintained, its resulting cellular structure, and its role in defining the transition between regimes.

We find a clear link between the DR and MC, with multi-cell MC profiles in the solar-like regime and single-cell MC profiles in the anti-solar regime.  We attribute this to the convective angular momentum transport (i.e.\ the zonal component of the convective Reynolds stress, ${\bf F}_{RS}$), which both establishes the DR contrast $\Delta \Omega$ and induces meridional flows through the mechanism of gyroscopic pumping.  In the solar-like regime, the positive $\Delta \Omega$ is maintained by a cylindrically outward transport (${\bf F}_{RS}$ directed away from the rotation axis) at low latitudes.  This induces a multi-cell MC profile outside the tangent cylinder, with two or more cells spanning the convection zone (CZ) at low latitudes, approximately aligned with the rotation axis.  At high latitudes radially inward angular momentum transport establishes a single circulation cell, with poleward flow in the upper CZ and equatorward flow in the lower CZ.

As the rotational influence decreases, the qualitative nature of ${\bf F}_{RS}$ changes.  We demonstrate this change by introducing a transport potential $\chi$ that highlights the orientation of ${\bf F}_{RS}$.   In particular, the high-latitude region of inward angular momentum transport extends to progressively lower latitudes as $Ro$ is increased, while the low-latitude transport becomes equatorward.  This establishes a single-celled MC profile throughout the CZ which is enhanced by baroclinic forcing.  Poleward angular momentum transport by this induced circulation is mainly responsible for reversing the sign of $\Delta \Omega$ and pushing the system into the anti-solar regime.  

Thus, though ${\bf F}_{RS}$ ultimately regulates the transition between mean flow regimes, it is the angular momentum transport by the MC, and not the Reynolds stress itself, that leads to the spin-up of the polar regions in anti-solar states.  This implies that additional factors impacting the efficiency of angular momentum transport by the MC, outside the purview of the simulations presented here, may play an important role in determining whether a particular star may attain an anti-solar state.  These factors include magnetism and subtle boundary layer influences involving the tachocline and near-surface shear layer.   The dynamics near the transition are subtle.  This is clear from the transitional case B0.5, which exhibits characteristics from both regimes, namely a solar-like DR and a single-celled MC.  Though we did not investigate it explicitly here, we expect our simulations (and perhaps real stars) may also exhibit bistability near the transition as reported by \cite{gasti14} and \cite{kapyl14}.

We emphasize that ${\bf F}_{RS}$ is the dominant factor that determines the MC profiles in our simulations through the mechanism of gyroscopic pumping.  This mechanism can operate even if thermal wind balance (TWB), eq.\ (\ref{eq:twb}) is satisfied everywhere.  Thus, it is notably different than in mean-field models where meridional flows are established by deviations from TWB due to meridional Reynolds or Maxwell stresses \citep{kitch12,dikpa14}.  These deviations are often localized in the boundary layers, which can have a disproportionate influence on the global MC profile.  Real stars likely have significant contributions from both mechanisms, further increasing the challenge of predicting the MC profiles for individual stars.  This is particularly the case for our own Sun, which may lie close to the transition between mean flow regimes.  Indeed, the shallow return flow inferred from recent helioseismic inversions and photospheric feature tracking may be a boundary-layer effect, reflecting the penetration depth of surface-driven convective plumes (sec.\ \ref{sec:nssl}).  

Nevertheless, the results presented here suggest that at the extremes of the rotation spectrum, the convective angular momentum transport is likely the dominant driver of DR and meridional circulation.  While the precise location of the transitional $Ro$ may depend on a number of factors, solar-like stars rotating significantly faster than the Sun should thus have cylindrical, solar-like DR profiles ($\Delta \Omega > 0$) 
and multi-celled MC profiles while those rotating significantly slower than the Sun should
have anti-solar DR profiles ($\Delta \Omega < 0$) and single-celled MC profiles.

%MMMMMMMMMMMMMMMMMMMMMMMMMMMMMMMMMMMMMMMMMMMMMMMMMM

\acknowledgements
This work is supported by NASA grants NNH09AK14I (Heliophysics SR\&T), NNX10AB81G, NNX09AB04G, NNX14AC05G, NNX13AG18G,
  and NNX08AI57G 
(Heliophysics Theory Program).  Additionally, Featherstone wishes to thank the High Altitude Observatory, whose gracious support and hospitality made this work possible.  The National Center for Atmospheric Research is 
sponsored by the National Science Foundation.  All simulations presented here were run on the Pleiades supercomputer at NASA Ames.

%\bibliographystyle{apj}
%\bibliography{apj-jour,mypapers,dynamo,helioseismology,observations,mean_field_hydro,convection_sphere,tachocline,convection_plane,general,stars}

\end{document}